\documentclass{llncs}

\usepackage[usenames,dvipsnames]{color}
\usepackage{graphicx}
\usepackage{subfigure}
\usepackage{algorithmic}
\usepackage{algorithm}
\usepackage{multirow}
\usepackage{cases}
\usepackage{hyperref}

\DeclareGraphicsExtensions{.pdf}

\newcommand{\secref}[1]{Section~\ref{sec_#1}}

\newcommand{\figref}[1]{Fig.~\ref{fig_#1}}
\newcommand{\tblref}[1]{Table~\ref{tbl_#1}}
\newcommand{\eqref}[1]{Eq.~(\ref{eq_#1})}
\newcommand{\algref}[1]{Alg.~\ref{alg_#1}}

\newcommand{\neigh}[0]{\mathrm{\Gamma}}
\newcommand{\cond}[0]{\mathrm{\Phi}}
\newcommand{\argmax}[0]{\arg\!\max}

\newcommand{\mpa}[0]{MPA}
\newcommand{\mpad}[0]{MPA(D)}
\newcommand{\mpap}[0]{MPA(P)}
\newcommand{\dmpa}[0]{MPA-D}
\newcommand{\bmpa}[0]{MPA-B}
\newcommand{\gpa}[0]{GPA}
\newcommand{\emmm}[0]{MM(EM)}
\newcommand{\gmo}[0]{MO(G)}
\newcommand{\km}[0]{KM}

\definecolor{brown}{RGB}{148, 81, 0}
\definecolor{blue}{RGB}{0, 122, 169}
\definecolor{green}{RGB}{77, 143, 42}
\definecolor{yellow}{RGB}{146, 144, 0}
\definecolor{lightgray}{RGB}{121, 121, 121}
\definecolor{darkgray}{RGB}{66, 66, 66}

\begin{document}

%	-----
% 	TITLE
%	-----

\title{Generalized network community detection}
\titlerunning{Generalized network community detection}
\author{Lovro \v{S}ubelj \and Marko Bajec}
\authorrunning{Lovro \v{S}ubelj \and Marko Bajec}
\institute{University of Ljubljana, Faculty of Computer and Information Science,\\
SI-1001 Ljubljana, Slovenia,\\
\email{[name.surname]@fri.uni-lj.si}}

\maketitle

%	--------
% 	ABSTRACT
%	--------

\begin{abstract}
Community structure is largely regarded as an intrinsic property of complex real-world networks. However, recent studies reveal that networks comprise even more sophisticated modules than classical cohesive communities. More precisely, real-world networks can also be naturally partitioned according to common patterns of connections between the nodes. Recently, a propagation based algorithm has been proposed for the detection of arbitrary network modules. We here advance the latter with a more adequate community modeling based on network clustering. The resulting algorithm is evaluated on various synthetic benchmark networks and random graphs. It is shown to be comparable to current state-of-the-art algorithms, however, in contrast to other approaches, it does not require some prior knowledge of the true community structure. To demonstrate its generality, we further employ the proposed algorithm for community detection in different unipartite and bipartite real-world networks, for generalized community detection and also predictive data clustering.
\keywords{link-density community, link-pattern community, propagation, community detection, data clustering}
\end{abstract}

%	--------	--------	   -----	--------	--------
% 					   PAPER
%	--------	--------	   -----	--------	--------

\section{\label{sec_intro}Introduction}
Over a decade of research in network analysis has revealed a number of common properties of complex real-world networks~\cite{WS98,FFF99}. \textit{Community structure}~\cite{GN02}---the occurrence of cohesive modules of nodes---is of particular interest as it provides an insight into not only structural organization but also functional behavior of various real-world systems~\cite{PDFV05,ADP06}. The analysis of communities has thus been the focus of many recent endeavors~\cite{For10,POM09}, while community structure analysis is also considered as one of the most prominent areas of network science~\cite{DDDA05,POM09}. 

\begin{figure}[t]
\centering
\subfigure[\label{fig_comms_ZKC}Zachary's karate network~\cite{Zac77}]{
\includegraphics[width=0.375\textwidth]{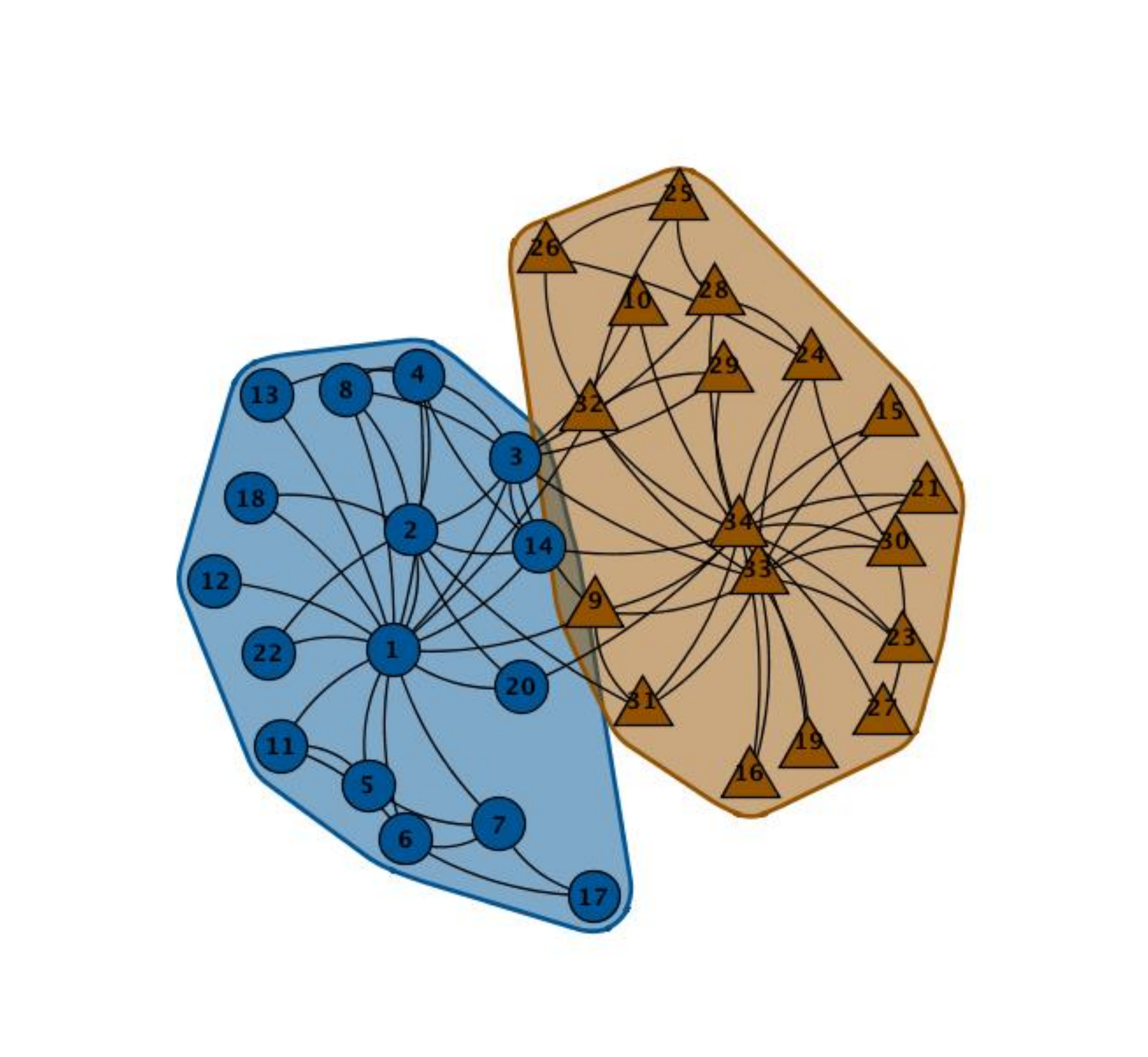}}
\subfigure[\label{fig_comms_SWC}Davis's women network~\cite{DGG41}]{
\includegraphics[width=0.35\textwidth]{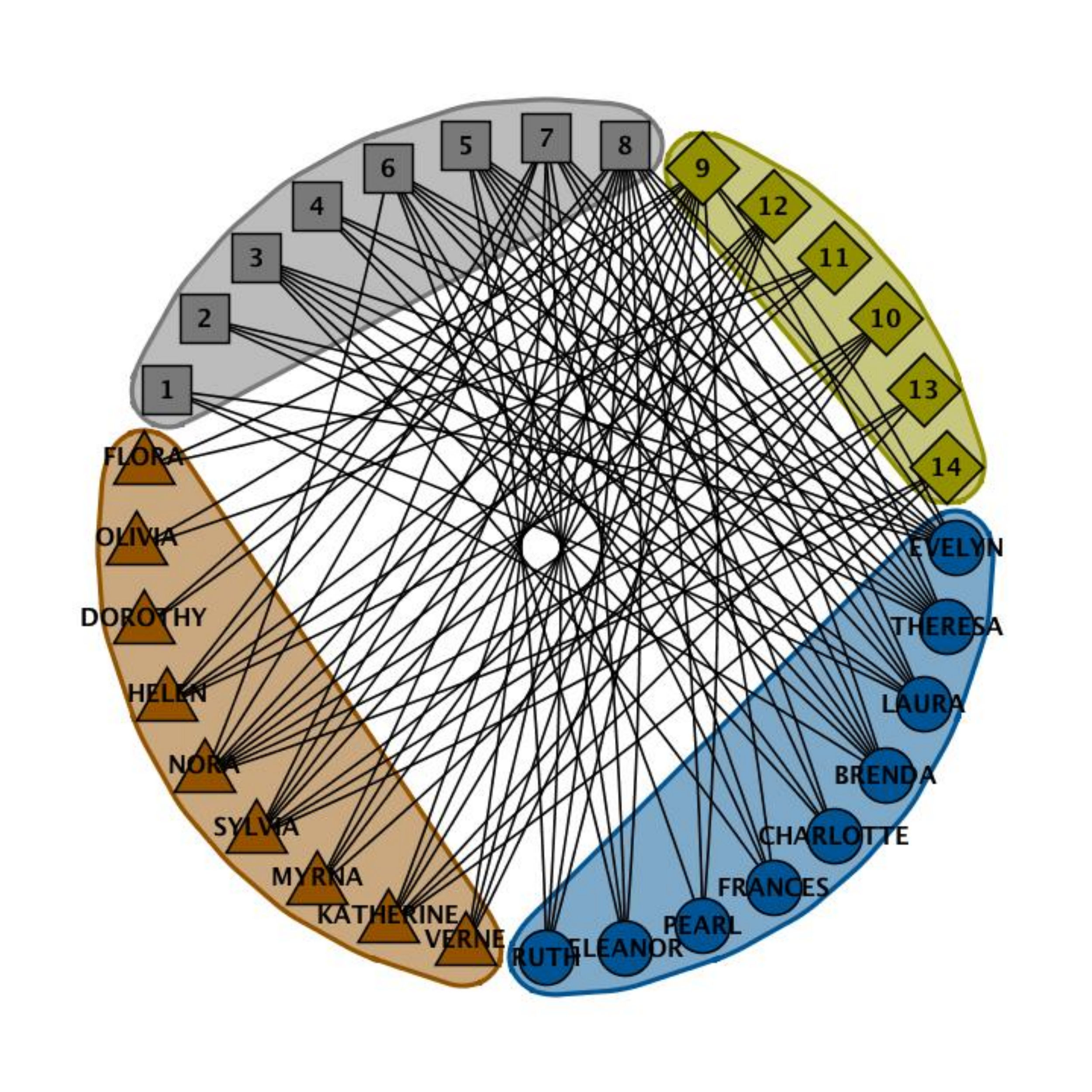}}
\caption{\label{fig_comms}Comparison between (a)~link-density and (b)~link-pattern communities.}
\end{figure}

However, most of the past work was constrained to communities characterized by higher density of links---\textit{link-density communities}~\cite{GN02} (\figref{comms_ZKC}). In contrast to the latter, recent studies reveal that networks comprise even more sophisticated modules than classical cohesive communities~\cite{NL07,ABFX08,PSR10,SB11c}. In particular, real-world networks can also be naturally partitioned according to common patterns of connections among nodes---into \textit{link-pattern communities}~\cite{LXZY07,NL07} (\figref{comms_SWC}). Link-pattern communities can in fact be related to relevant functional roles in various complex systems~\cite{PSR10,SB11c}, moreover, they also provide a further comprehension of real-world network structure that is obscure under classical frameworks. Note that link-density communities could be seen as a special case of link-pattern communities, although several fundamental differences exist~\cite{SB11c}. In particular, link-pattern communities do not correspond to densely connected groups of nodes, while generally do not even feature connectedness. The latter actually implies low transitivity---clustering coefficient~\cite{WS98}---for the nodes in link-pattern communities, which contradicts with small-world phenomena~\cite{WS98}. However, recent work suggests that best link-pattern communities might indeed emerge in parts of networks that exhibit low values of clustering (e.g., technological networks), where small-world property does not generally hold~\cite{SB11f}.

Recently, \v{S}ubelj~and~Bajec~\cite{SB11c} have proposed a general propagation algorithm that can reveal arbitrary network modules ranging from link-density to link-pattern communities. Their algorithm does not require any prior knowledge of the true structure, though they introduce a community parameter that models the nature of each community according to the measure of network bottlenecks---conductance~\cite{Bol98}. We advance the latter by proposing a more adequate modeling strategy based on node clustering coefficient~\cite{WS98}. The resulting algorithm is evaluated on various synthetic benchmark networks with planted partition, on random graphs and also resolution limit examples. It is shown to be comparable to current state-of-the-art, whereas, the proposed strategy also greatly improves on the approach of~\v{S}ubelj~and~Bajec~\cite{SB11c} (on these networks). Furthermore, to demonstrate its generality, we also employ the algorithm for community detection in different unipartite and bipartite real-world networks, for generalized community detection and predictive data clustering.

The rest of the paper is structured as follows. In~\secref{rw} we briefly review relevant related work, with emphasis on the community detection literature. \secref{alg} introduces the proposed algorithm, while the empirical evaluation with formal discussion is done in~\secref{eval}. The performance on various real-world examples is presented in~\secref{rwe}, and conclusions are made in~\secref{conc}.

%	--------	--------	--------

\section{\label{sec_rw}Related work}
Despite the wealth of the literature on classical communities in recent years~\cite{For10}, only a small number of authors have considered more general link-pattern communities. Nevertheless, authors have recently proposed different algorithms based on stochastic blockmodels~\cite{ABFX08,HW08}, mixture models~\cite{NL07,SC10a}, model selection~\cite{LXZY07,PSR10}, data clustering~\cite{LKC10} and other~\cite{PMB10}. However, in contrast to the propagation algorithm proposed in this paper, and that in~\cite{SB11c}, all other approaches require some prior knowledge of the true structure (e.g., the number of communities). The latter indeed seriously limits their use in practice. Note that authors have also analyzed vertex similarity based on common patterns of connections~\cite{BGHSD04,LHN06}---commonly referred to as \textit{structural equivalence}---whereas, some of the research on classical communities also apply for link-pattern counterparts~\cite{GA05,RB07}. 

It ought to be mentioned that link-pattern communities are known as \textit{block\-models}~\cite{WBB76} in social networks literature. These have been extensively studied in the past, however, their main focus and employed formulation differs from ours.

%	--------	--------	--------

\section{\label{sec_alg}Model-based propagation}
Let the network be represented by an undirected graph $G(N,L)$, where $N$ is the set of nodes of the graph and $L$ is the set of its links (edges). Furthermore, let $w_{nm}$ be the weight of the link between nodes $n,m\in N$. Moreover, let $c_n$ denote the community (label) of node $n\in N$, and let $\neigh_n$ be the set of its neighbors.

The proposed model-based propagation algorithm is, as the algorithm in~\cite{SB11c}, based on the label propagation principle of~Raghavan~et~al.~\cite{RAK07}. In the following, we thus first introduce the latter.

\paragraph{Label Propagation.}Label propagation algorithm~\cite{RAK07} (LPA) reveals link-density communities by exploiting the following procedure. First, each node $n\in N$ is labeled with a unique label (i.e., $c_n=l_n$). Then, at each iteration, the node adopts the label shared by most of its neighbors (with respect to link weights).~Hence,
\begin{eqnarray}
\label{eq_lpa}
c_n & = & \argmax_l\sum_{m\in\neigh_n^l}w_{nm},
\end{eqnarray}
where $\neigh_n^l$ is the set of neighbors of node $n$ that share label $l$ (ties are broken uniformly at random\footnote{When node's current label is among most frequent, the node retains its label.}). Due to the existence of many intra-community links, relative to the number of inter-community links, cohesive modules of nodes form a consensus on some label after a few iterations. Thus, when the algorithm converges---a local equilibrium is reached---disconnected sets of nodes sharing the same label are classified into the same community. Due to extremely fast structural inference of label propagation, the algorithm exhibits near linear complexity and can easily scale to networks with millions of nodes and links~\cite{SB11d,RAK07}.

Note that, to address issues with oscillations of labels in some networks (e.g., bipartite networks), label updates in~\eqref{lpa} occur in a random order~\cite{RAK07}.

\paragraph{General Propagation.}\v{S}ubelj~and~Bajec~\cite{SB11c} have argued that label propagation cannot be directly applied for the detection of link-pattern communities, as the bare nature of propagation requires connected (and cohesive) groups of nodes. However, when one considers second order neighborhoods, and propagates labels through nodes' neighbors, link-pattern communities indeed correspond to cohesive modules of nodes (see~\figref{comms_SWC}). Based on the above they have proposed general propagation algorithm~\cite{SB11c} (\gpa) that is presented in the following.

Let $\delta_c$ be a community parameter that models the nature of community $c$, $\delta_c\in[0,1]$. Assume $\delta_c$ equals $1$ and $0$ for link-density and link-pattern communities, respectively (to be properly defined later). Label propagation in~\eqref{lpa} is then advanced into a general community detection algorithm as
\begin{eqnarray}
\label{eq_mpa}
c_n & = & \argmax_l\left(\delta_l\sum_{m\in\neigh_n^l}w_{nm}b_md_m + (1-\delta_l)\sum_{m\in\neigh_s^l\backslash\neigh_n|s\in\neigh_n}w_{nm}^sb_m\tilde{d}_m\right)
\end{eqnarray}
where $w_{nm}^s=\frac{w_{ns}w_{sm}}{s_s}$ and $s_n$ is the strength of node $n\in N$ (i.e., $s_n=\sum_{m\in\neigh_n}w_{nm}$). In the case of link-density communities (left-hand side of~\eqref{mpa}), the labels are propagated among the neighbors as before, whereas, in the case of link-pattern communities (right-hand side of~\eqref{mpa}), the labels are propagated through nodes' neighbors---between the nodes at distance two. Thus, the algorithm can indeed reveal either link-density or link-pattern communities, or different mixtures of both, when they are clearly depicted in the network's topology. (Note that in~\cite{SB11c} the algorithm was presented for unweighted networks.)

Node balancers $b_n$~\cite{SB11b} and diffusion values $d_n$, $\tilde{d}_n$~\cite{SB11d,SB11c} in~\eqref{mpa} improve the algorithm's stability and accuracy, respectively. More precisely, random label update orders (see above) severely hamper the robustness of the approach, and consequently also the stability of the identified community structure~\cite{TK08}. In particular, nodes that are updated at the beginning exhibit higher propagation preferences than those that are updated towards the end~\cite{SB11b}. Thus, balancers $b_n$ are utilized to counteract for the randomness introduced by update orders---lower and higher preferences are given to the nodes updated first and last, respectively.

Let $i_n$ denote a normalized position of node $n\in N$ in some random order, $i_n\in(0,1]$. Then, node balancers are set according to
\begin{eqnarray}
\label{eq_b}
b_n & = & \frac{1}{1+e^{-\mu(i_n-\lambda)}},
\end{eqnarray}
where $\lambda$ and $\mu$ are parameters of the algorithm. Intuitively, we fix $\lambda$ to $\frac{1}{2}$, while $\mu$ is set to $2$ based on some preliminary experiments (see~\secref{eval}). Node balancers can also be modeled with a linear function as $b_n=i_n$, however, introduction of the above parameters allows for a distinct control over the algorithm. In particular, analysis in~\secref{eval} reveals that increasing $\mu$ improves the stability of the algorithm, although the computational time thus also increases. Note also that setting $\mu$ to $0$ yields a classical label propagation where all $b_n$ are equal.

To further boost the community detection strength of the algorithm, defensive preservation of communities is employed through diffusion values $d_n$, $\tilde{d}_n$~\cite{SB11d,SB11c}. Here higher diffusion values---propagation preferences---are given to core nodes of each (current) community, while lower values are given to their border nodes.  The latter results in an immense ability of detecting communities, even when they are only weakly depicted in the network's topology~\cite{SB11d}. At each iteration, diffusion values are estimated by means of a random walker utilized on each (current) community. Hence,
\begin{eqnarray}
d_n = \sum_{m\in\neigh_n^{c_n}}d_m/k_m^{c_n}
\end{eqnarray}
and
\begin{eqnarray}
\tilde{d}_n = \sum_{m\in\neigh_s^{c_n}\backslash\neigh_n|s\in\neigh_n}\frac{\tilde{d}_m}{\sum_{s\in\neigh_m}k_{s}^{c_n}},
\end{eqnarray}
where $k_{n}^{c_n}$ is the intra-community degree of node $n\in N$ (all $d_n$, $\tilde{d}_n$ are initialized to $\frac{1}{|N|}$). Besides deriving an estimate of the core and border of each community, the main rationale here is to formulate propagation---diffusion---within each community, to estimate the current state of label propagation, and then to adequately alter the dynamics of the process. Analysis in~\secref{eval} reveals that defensive preservation of communities significantly improves the detection strength of the algorithm, while for further discussion and analysis see~\cite{SB11d}.

Despite the discussion above, the core of the algorithm is in fact represented by a community modeling strategy implemented through parameters $\delta_c$. \v{S}ubelj~and~Bajec~\cite{SB11c} have proposed to measure the conductance~\cite{Bol98} of each community, to determine whether it better conforms with link-density or link-pattern regime. Conductance~$\cond(c)$ of community $c$ is defined as a relative size of the corresponding network cut---ratio of inter-community links---thus it is a measure of network bottlenecks. Hence, at each iteration, they simply set $\delta_c=1-\cond(c)$, while all $\delta_c$ are initialized to $\frac{1}{2}$. The main weakness of their strategy is that each community is considered independently of other. Thus, in the following, we propose a more adequate~community modeling strategy based on the properties of complex real-world networks.

% For further discussion on general propagation algorithm see~\cite{SB11c,SB11b,SB11d}.

\paragraph{Model-based Propagation.}Community modeling strategy of~\v{S}ubelj~and~Bajec~\cite{SB11c} considers merely the nature of each respective community, whereas all other communities are disregarded. Although no proper empirical study exists, in an ideal case, link-pattern communities would link to other link-pattern communities rather than to other link-density communities. The latter follows from the fact that the concerned links would else obviously decrease the quality of the respective link-density community---make it a link-pattern community. Thus, we propose a community model based on the hypothesis that the neighbors' communities should be of the same type---either link-density or link-pattern---as the concerned node's community. Hence,
\begin{eqnarray}
\label{eq_delta}
\delta_c & = & \frac{1}{|N^c|}\sum_{m\in\neigh_n|n\in N^c}\frac{\delta_{c_m}}{k_n},
\end{eqnarray}
where $k_n$ is the degree of node $n\in N$ and $N^c$ is the set of nodes in community~$c$. 

We also argue that an adequate initialization of community parameters $\delta_c$ is of vital importance (exact results are omitted). Otherwise, the algorithm can easily get trapped in some local stable---probably suboptimal---fixed point that is hard to escape from. However, \eqref{delta} cannot be directly employed at the beginning, as all nodes still reside in their own communities. We thus refine the above hypothesis such that the node's neighbors should not only reside in the same type of the community, but in the same respective community. The latter immediately implies that the neighbors of the nodes in link-density communities should also link to each other, whereas the opposite holds for the nodes in link-pattern communities. Hence, for each node $n\in N$, one could initially set $\delta_{c_n}$ to $C_n$, where $C_n$ is a node clustering coefficient~\cite{WS98} defined as the probability that two neighbors of node $n$ also link to each other---network transitivity.  It ought to be mentioned that recent work suggests that transitivity---rather than homophily---gives rise to the modular structure in real-world networks~\cite{FFGP10}.

However, consider a node with very high degree---a hub node. Hubs commonly appear in link-density communities~\cite{GA05}, still, due to a large number of links, they would only rarely experience high values of clustering coefficient (the opposite would in fact imply a large clique). Also, as most networks are disassortative by degree~\cite{New02}, hubs tend to link to low degree nodes that cannot provide for high clustering of the hub node~\cite{SV05}. Indeed, in many real-world networks node clustering coefficient roughly follows $C_n\sim k_n^{-1}$~\cite{VPV02,RB03,SV05}, where $k_n$ is the degree of node $n\in N$. Hence, we model initial communities as (assume $C_n>0$)
\begin{subnumcases}{\label{eq_init}\delta_{c_n} = }
1 & for $C_n>\alpha k^{-1}_n+\beta$,\\
\rho & otherwise,
\end{subnumcases}
where $\alpha$ and $\beta$ are estimated from the network using ordinary least squares, and $\rho$ is a parameter. We set $\rho$ to $\frac{1}{4}$ based on some preliminary experiments.

\eqref{init}~and~\eqref{delta} define the proposed model-based propagation algorithm (\mpa), which is else (almost) identical to the algorithm in~\cite{SB11c} (see~\algref{mpa}). However, the evaluation on synthetic and real-world networks in~\secref{eval}~and~\secref{rwe}, respectively, reveals that the proposed approach significantly outperforms that~in~\cite{SB11c}. For a thorough evaluation, we also analyze two variations of the basic approach that fix all community parameters $\delta_c$ to either $1$ or $0$. The approaches thus result in a fully link-density or link-pattern community detection algorithms, and are denoted \mpad~and~\mpap, respectively.

\begin{algorithm}[t]
\algsetup{indent=1em}
\caption{\label{alg_mpa}Model-based propagation algorithm (MPA)}
\begin{algorithmic}[0]
\REQUIRE Graph $G(N,L)$ and parameters $\lambda$, $\mu$, $\rho$
\ENSURE Communities $C$
\STATE \COMMENT{Community initialization.}
\FOR{$n\in N$} 
	\STATE $c_n\gets l_n$\COMMENT{Unique label.}
	\STATE $\delta_{c_n}\gets$ \COMMENT{Model according to~\eqref{init}.}
	\STATE $d_n, \tilde{d}_n\gets 1/|N|$
\ENDFOR
\STATE \COMMENT{Model-based propagation.}
\WHILE{\NOT \textit{converged}}
	\STATE \textbf{shuffle}$(N)$
	\FOR{$n\in N$} 	
		\STATE \COMMENT{General propagation.}
		\STATE $b_n\gets 1/(1+e^{-\mu(i_n-\lambda)})$
		\STATE $c_n\gets \argmax_l\left(\delta_l\sum_{m\in\neigh_n^l}w_{nm}b_md_m + (1-\delta_l)\sum_{m\in\neigh_s^l\backslash\neigh_n|s\in\neigh_n}w_{nm}^sb_m\tilde{d}_m\right)$
		\STATE \COMMENT{Re-estimation.}
		\STATE $d_n\gets \sum_{m\in\neigh_n^{c_n}}d_m/k_m^{c_n}$ and $\tilde{d}_n\gets \sum_{m\in\neigh_s^{c_n}|s\in\neigh_n}\tilde{d}_m/\left.\sum_{s\in\neigh_m}k_{s}^{c_n}\right.$
	\ENDFOR
	\FOR{$c\in C$} 	
		\STATE \COMMENT{Community modeling.}
		\STATE $\delta_c\gets 1/|N^c|\sum_{m\in\neigh_n|n\in N^c}\delta_{c_m}/k_n$ \COMMENT{Omitted on first iteration.}
	\ENDFOR
\ENDWHILE
\RETURN $C$
\end{algorithmic}
\end{algorithm}

%	--------	--------	--------

\section{\label{sec_eval}Evaluation and discussion}
In the following we evaluate the proposed algorithm on different synthetic benchmark networks with planted partition, and also on random networks. The results are assessed in terms of three different measures of community significance, borrowed from the field of information theory and community detection literature.

Let $\mathcal{C}$ be a partition extracted by an algorithm and let $\mathcal{P}$ be the known partition of the network (corresponding random variables are $C$ and $P$, respectively). First---normalized mutual information~\cite{DDDA05} (NMI)---has become a de facto standard in the recent literature. NMI of $\mathcal{C}$ and $\mathcal{P}$ is defined as $\frac{2I(C,P)}{H(C)+H(P)}$, where $I(C,P)$ is the mutual information, and $H(C)$, $H(P)$ and $H(C|P)$ are standard and conditional entropies. NMI of identical partitions equals $1$, and is $0$ for independent ones, $\mathrm{NMI}\in[0,1]$.
Second, we also consider normalized variation of information~\cite{Mei07,KLN08} (NVOI), which is a symmetric local measure that has the properties of a distance in the space of partitions. NVOI of $\mathcal{C}$ and $\mathcal{P}$ equals $\frac{H(C|P)+H(P|C)}{\log|N|}$, therefore, in contrast to NMI, lower values represent better correlation between partitions, $\mathrm{NVOI}\in[0,1]$. Last, for a better comprehension, we also adopt a more intuitive measure---fraction of correctly classified nodes~\cite{GN02} (FCC)---that is commonly adopted within community detection literature. The node is considered correctly classified, if it resides in the same community as at least one half of the nodes in its true community. Again, $\mathrm{FCC}\in[0,1]$.

Community detection algorithms introduced in~\secref{alg} are compared against a greedy agglomerative optimization~\cite{New04a,CNM04} of modularity~$Q$~\cite{NG04} (denoted \gmo) ---a classical link-density community detection algorithm---and a mixture model with expectation-maximization~\cite{DLR77} proposed by~Newman~and~Leicht~\cite{NL07} (denoted \emmm). The latter can detect arbitrary network modules and is currently among state-of-the-art approaches for generalized community detection~\cite{NL07,PSR10}. However, it demands the correct number of communities to be known ahead of time, which puts the algorithm in significant advantage compared to others~\cite{KN11a}. For simplicity, we limit the number of iterations to $100$ for all algorithms.

\begin{figure}[p]
\centering
\subfigure[Analysis subject to NMI]{
\includegraphics[width=0.475\textwidth]{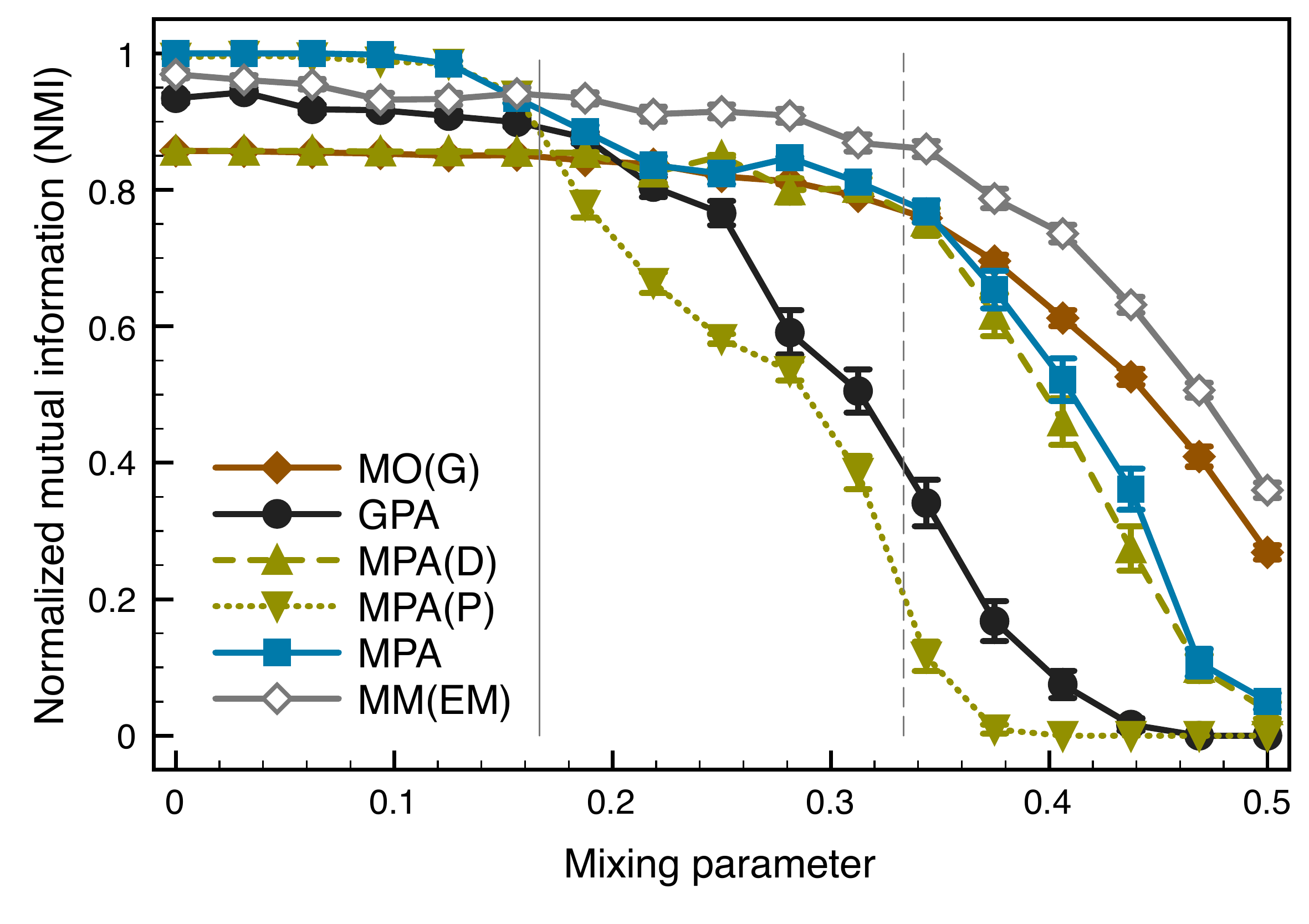}}
\subfigure[Analysis subject to NVOI]{
\includegraphics[width=0.475\textwidth]{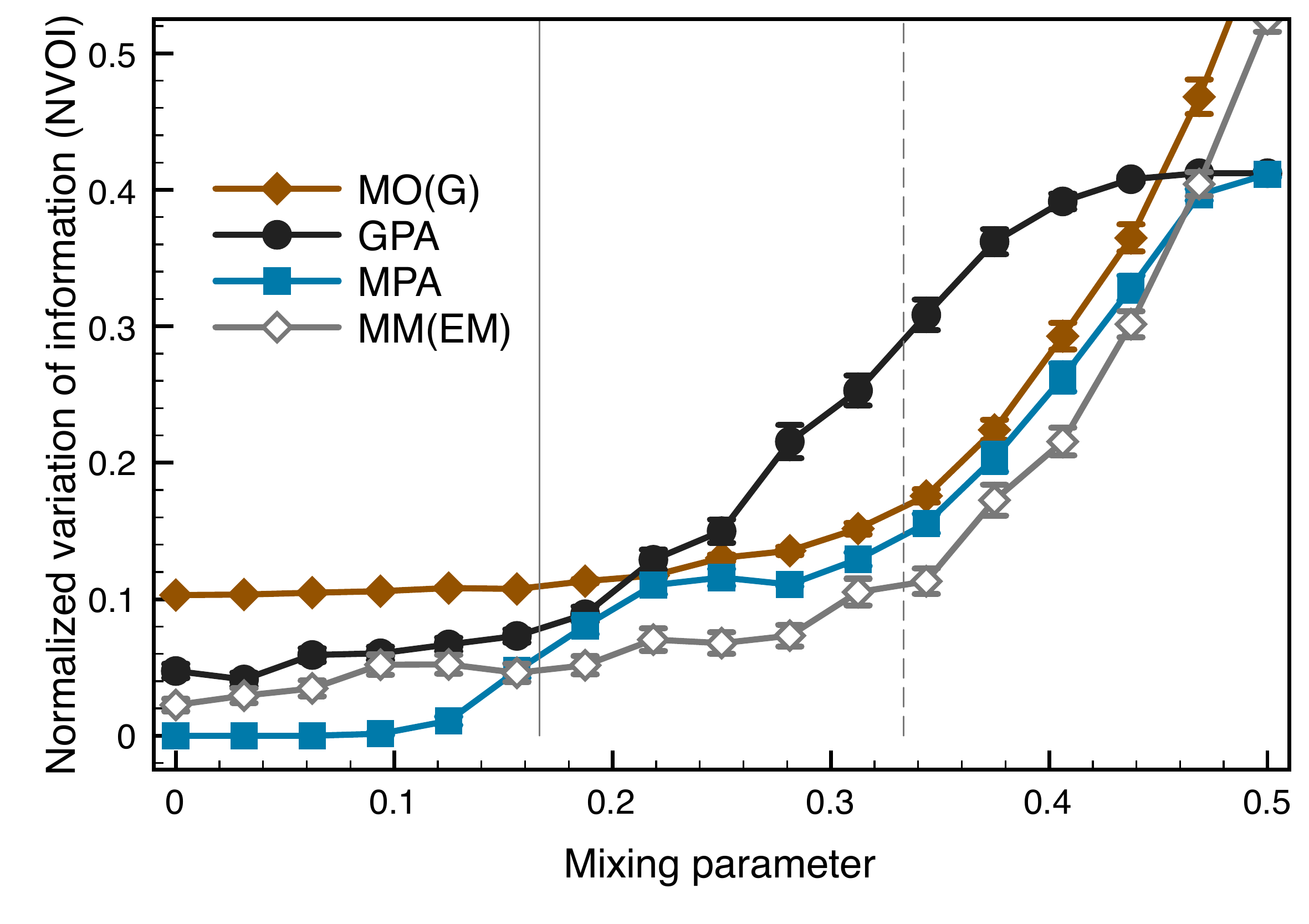}}
\caption{\label{fig_eval_GN2}Analysis on GN2 benchmark networks~\cite{PSR10}. The values are estimates over $100$ network realizations, while error bars show standard error of the mean.}
\end{figure}

\begin{figure}[p]
\centering
\subfigure[Analysis subject to NMI]{
\includegraphics[width=0.475\textwidth]{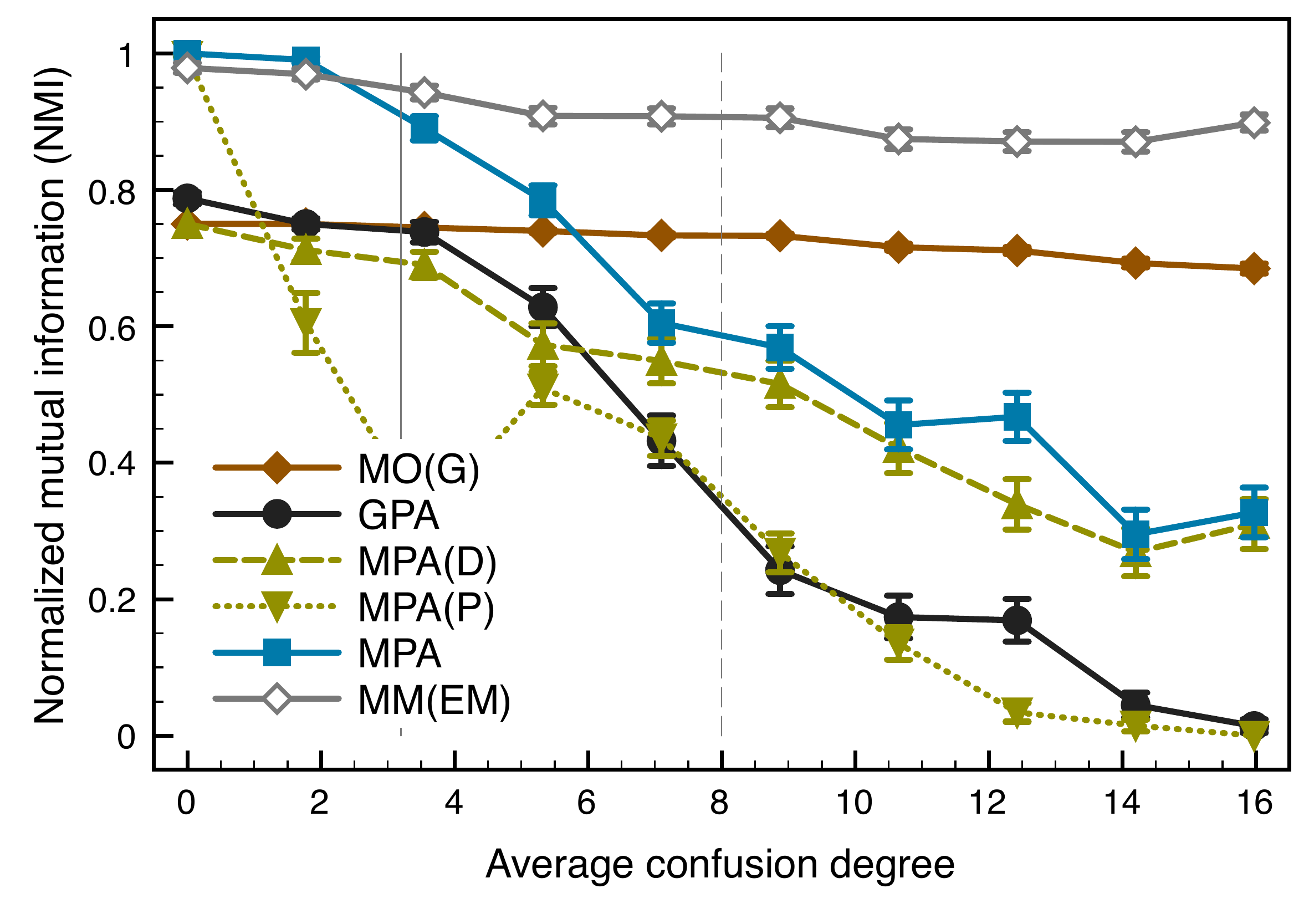}}
\subfigure[Analysis subject to NVOI]{
\includegraphics[width=0.475\textwidth]{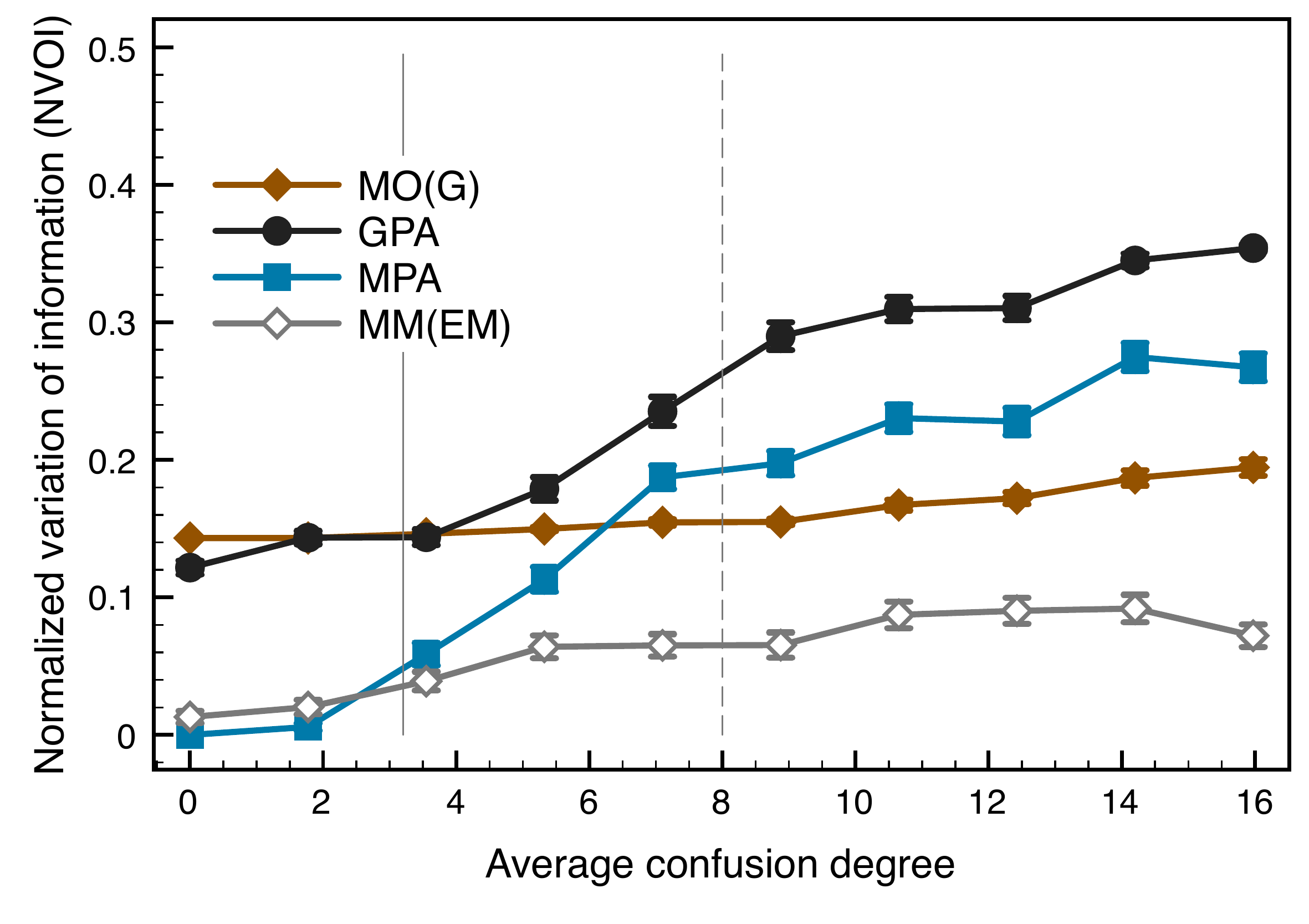}}
\caption{\label{fig_eval_SB}Analysis on SB benchmark networks. The values are estimates over $100$ network realizations, while error bars show standard error of the mean.}
\end{figure}

\paragraph{GN2 Benchmark.}The algorithms are first applied to a class of benchmark networks~\cite{PSR10} that is in fact a generalization of a classical benchmark proposed by~Girvan~and~Newman~\cite{GN02}. Networks comprise four communities of $32$ nodes, whereas, two communities correspond to classical link-density modules, while the other two form a bipartite structure of link-pattern communities. Average degree is fixed to $16$, while the community structure is controlled by a mixing parameter $\theta$, $\theta\in[0,1]$. When $\theta$ is $0$, all links are set according to the designed community structure, while for $\theta$ equal $1$, the networks are completely random.

The results are shown in~\figref{eval_GN2}. Observe that for small values of $\theta$ only \mpa~and \mpap~can accurately reveal the planted structure in these networks. However, when $\theta$ increases, the performance of \mpa~is similar to that of a classical community detection algorithm (e.g., \gmo~or \mpad). \emmm~can detect communities to some extent until $\theta\leq\frac{1}{3}$ (dashed lines in~Figs.~\ref{fig_eval_GN2},~\ref{fig_eval_SB})---when, for the nodes within link-density communities, there are twice as many links that conform with the planted structure than randomly placed links. Note also that twice as many links are needed to define a link-pattern community, compared to a respective link-density community, which would yield the same threshold at $\theta=\frac{1}{6}$ for these networks (solid lines in~Figs.~\ref{fig_eval_GN2},~\ref{fig_eval_SB}). Thus, \mpa~accurately extracts planted link-density and link-pattern communities in these networks, as long as they are clearly depicted in the network's topology. Note also that community modeling strategy within \mpa~seems more adequate~than~that~of~\gpa.

\begin{figure}[t]
\centering
\subfigure[\label{fig_SB}SB benchmark network]{
\includegraphics[width=0.375\textwidth]{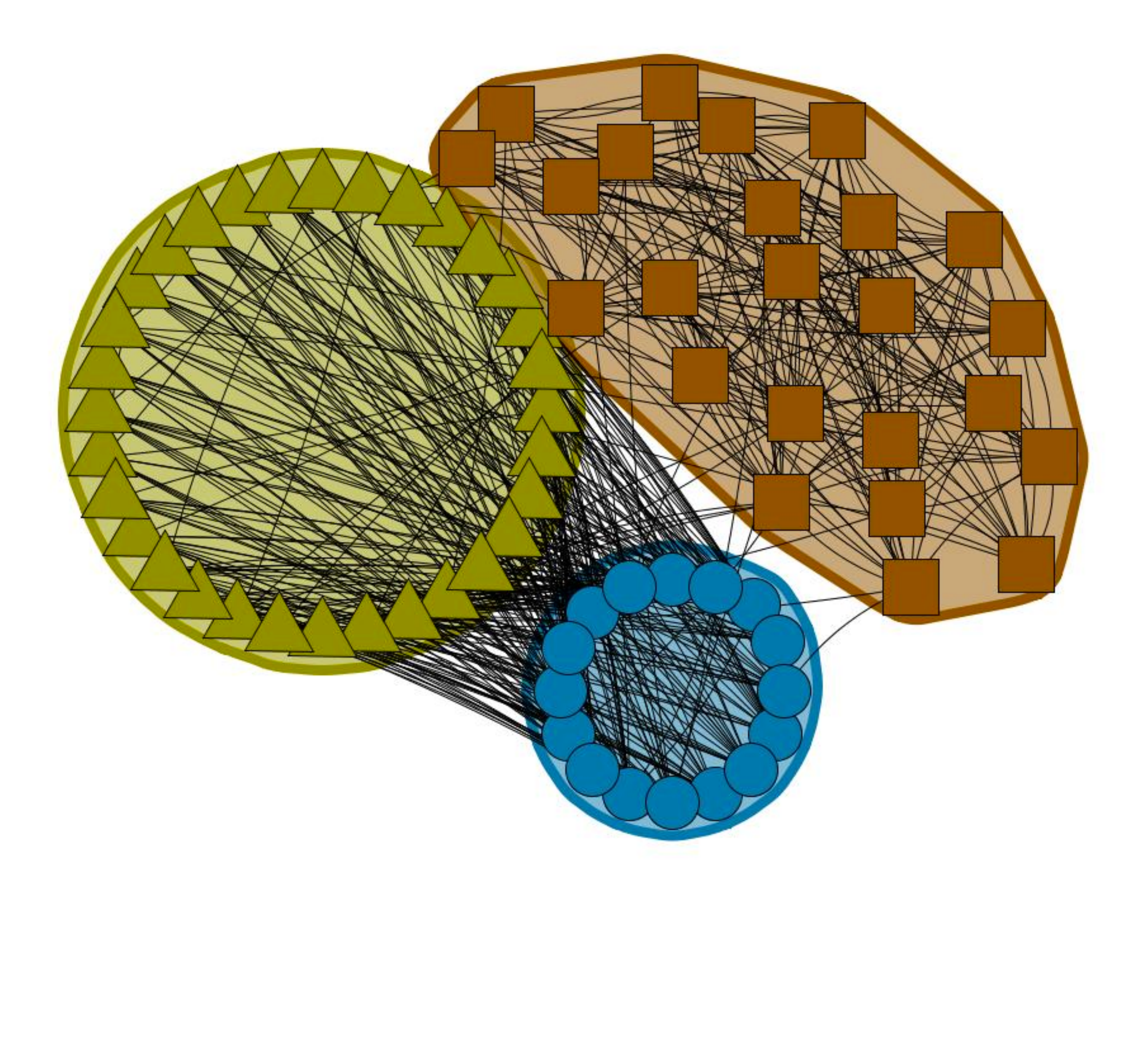}}
\subfigure[\label{fig_eval_SB_delta_vals}Analysis of community model in~\mpa]{
\includegraphics[width=0.475\textwidth]{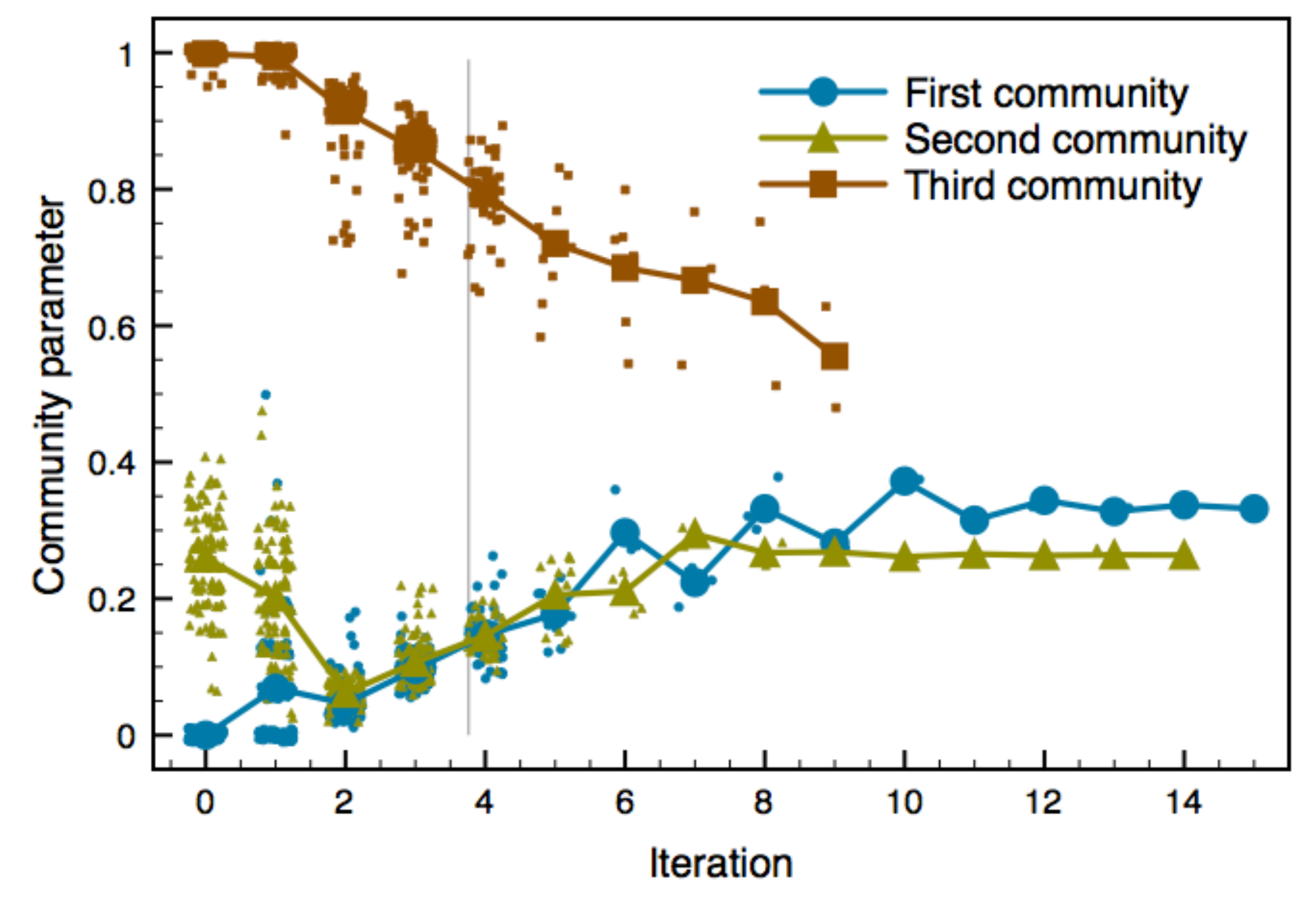}}
\caption{\label{fig_eval_SB_delta}Analysis of community modeling strategy of~\mpa~on SB benchmark networks. Node shapes represent planted network communities and are consistent among figures. The values are estimates over $100$ network realizations, while, for an adequate analysis, we set $\rho$ to $0$ and increase $\mu$ (\secref{alg}). See also text.} % $\mu$ is set to $8$.
\end{figure}

\paragraph{SB Benchmark.}GN2 benchmark provides a rather unrealistic testbed due to homogeneous degree and community size distributions. We address the latter by proposing a class of simple benchmark networks with heterogeneous community sizes. Networks comprise three communities of $16$, $32$ and $24$ nodes, respectively (see network in~\figref{SB}). The latter two again form a bipartite structure of link-pattern communities, while the third community corresponds to a classical cohesive module. Links are placed according to the designed community structure such that the average degree of the nodes in the first and third community is fixed to $16$. The latter implies an average degree of $8$ for the nodes in the second community. Furthermore, we also add some number of links uniformly at random for each node---denoted node confusion degree $\kappa$, $\kappa\geq 0$.

\begin{figure}[p]
\centering
\subfigure[Analysis subject to NMI]{
\includegraphics[width=0.475\textwidth]{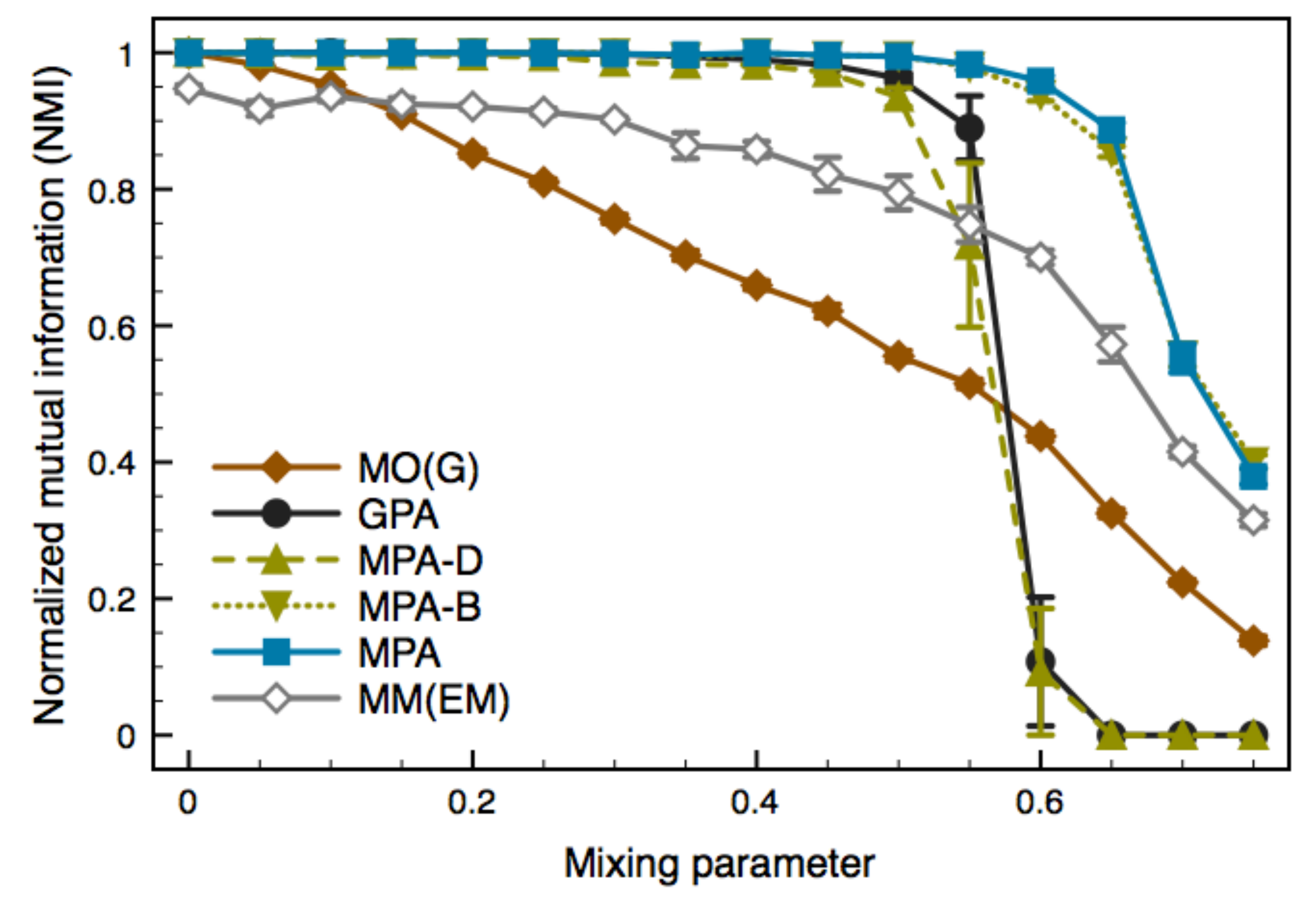}}
\subfigure[Analysis subject to NVOI]{
\includegraphics[width=0.475\textwidth]{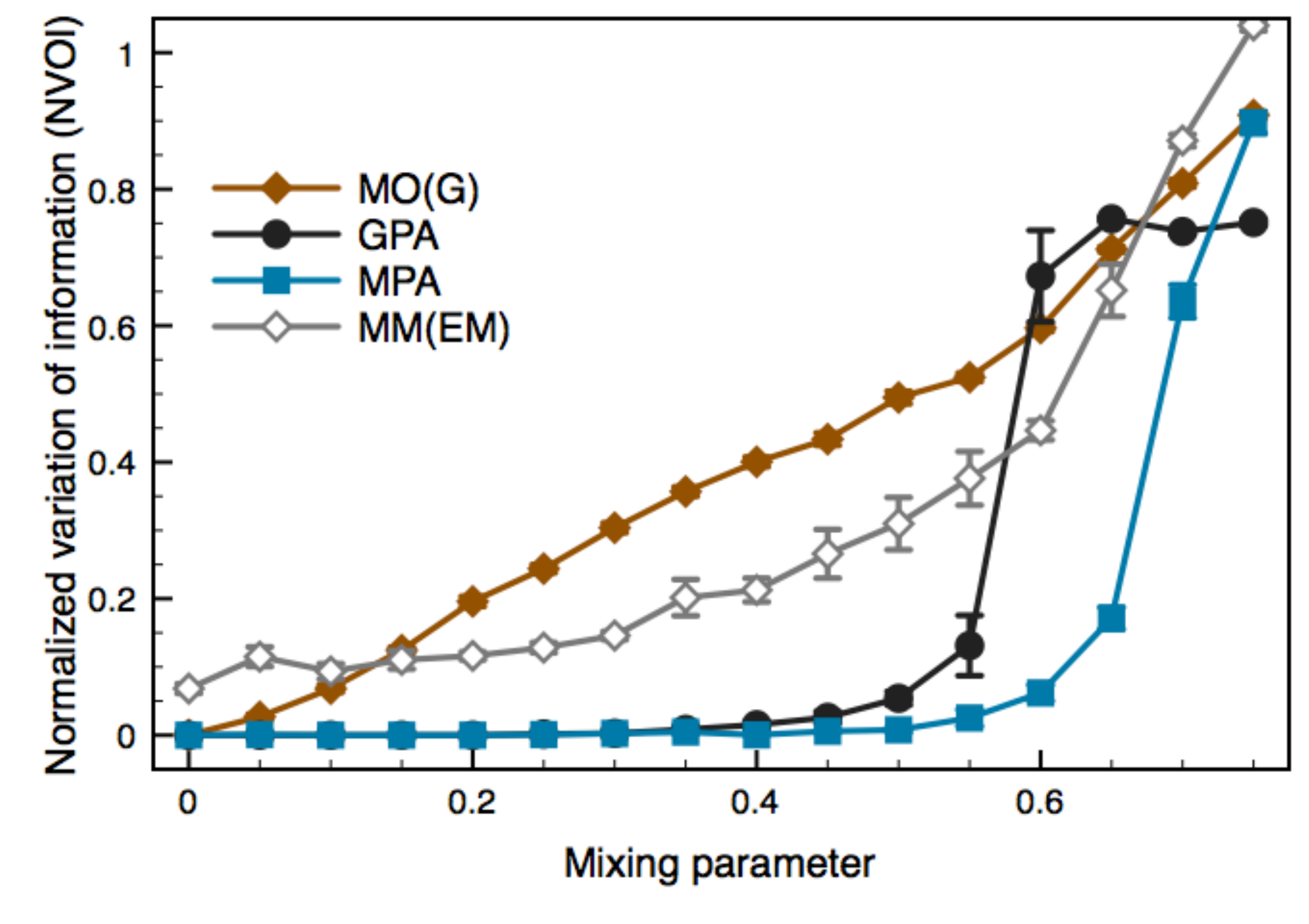}}
\caption{\label{fig_eval_LFR}Analysis on LFR benchmark networks. The values are estimates over $10$ network realizations, while error bars show standard error of the mean. To ensure convergence, $\mu$ is set to $\frac{1}{2}$.}
\end{figure}

The results appear in~\figref{eval_SB}. The performance of the algorithms is rather similar to that on GN2~benchmark (note different scales in~Figs.~\ref{fig_eval_GN2},~\ref{fig_eval_SB}). Only \mpa~can accurately reveal the planted structure for small values of~$\kappa$, while the model within~\gpa~again seems to fail. Observe that \emmm~can extract communities equally well, even when $\kappa$ equals $16$---only $\frac{1}{3}$ of the links for the nodes in the second community still agrees with the intrinsic structure, thus, the communities are only marginally defined. The latter clearly demonstrates that knowing an exact number of communities indeed presents a significant advantage.

\paragraph{LFR Benchmark.}To enable easier comparison with previous literature on community detection, we also apply the algorithms to a class of standard benchmark networks with scale-free degree and community size distributions proposed by Lancichinetti~et~al.~\cite{LFR08}. The size of the networks is set to $1000$, while community sizes range between $10$ and $50$ nodes. Note that all communities here correspond to a link-density regime. As before, the quality of the planted structure is controlled by a mixing parameter $\theta$, $\theta\in [0,1]$. For comparison, we also analyze two variations of \mpa~that do not employ either balanced propagation or defensive preservation of communities (denoted \dmpa~and \bmpa, respectively).

Results in~\figref{eval_LFR} show that \mpa~most accurately reveals the planted structures in these networks, while it also significantly outperforms the other generalized community detection algorithm \emmm. Observe also that defensive preservation of communities greatly improves the algorithm's community detection strength. Comparing the results with an analysis of over ten state-of-the-art approaches for classical community detection conducted in~\cite{LF09b}, we conclude that, at least on these networks, \mpa~performs similarily as the best algorithms analyzed there. These are hierarchical modularity optimization of Blondel~et~al.~\cite{BGLL08}, model selection technique of Rosvall~and~Bergstrom~\cite{RB08}, spectral algorithm proposed by Donetti~and~Mu\~noz~\cite{DM04} and multi-resolution spin model of Ronhovde~and~Nussinov~\cite{RN10}.

\begin{figure}[t]
\centering
\subfigure[\label{fig_HN_dend}Dendrogram]{
\includegraphics[width=0.25\textwidth]{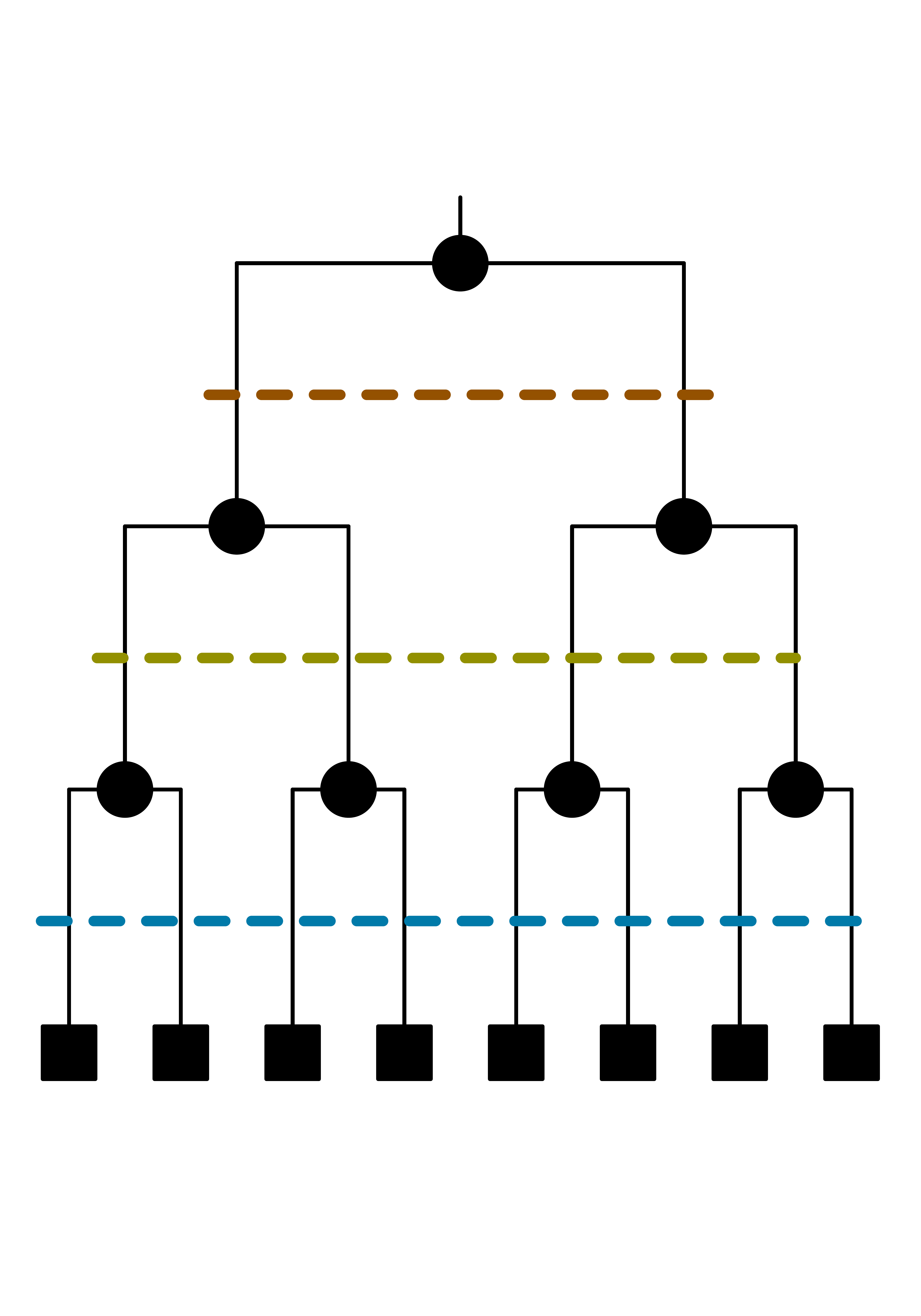}}
\subfigure[\label{fig_eval_HN_vals}Analysis subject to NMI---without fills]{
\includegraphics[width=0.50\textwidth]{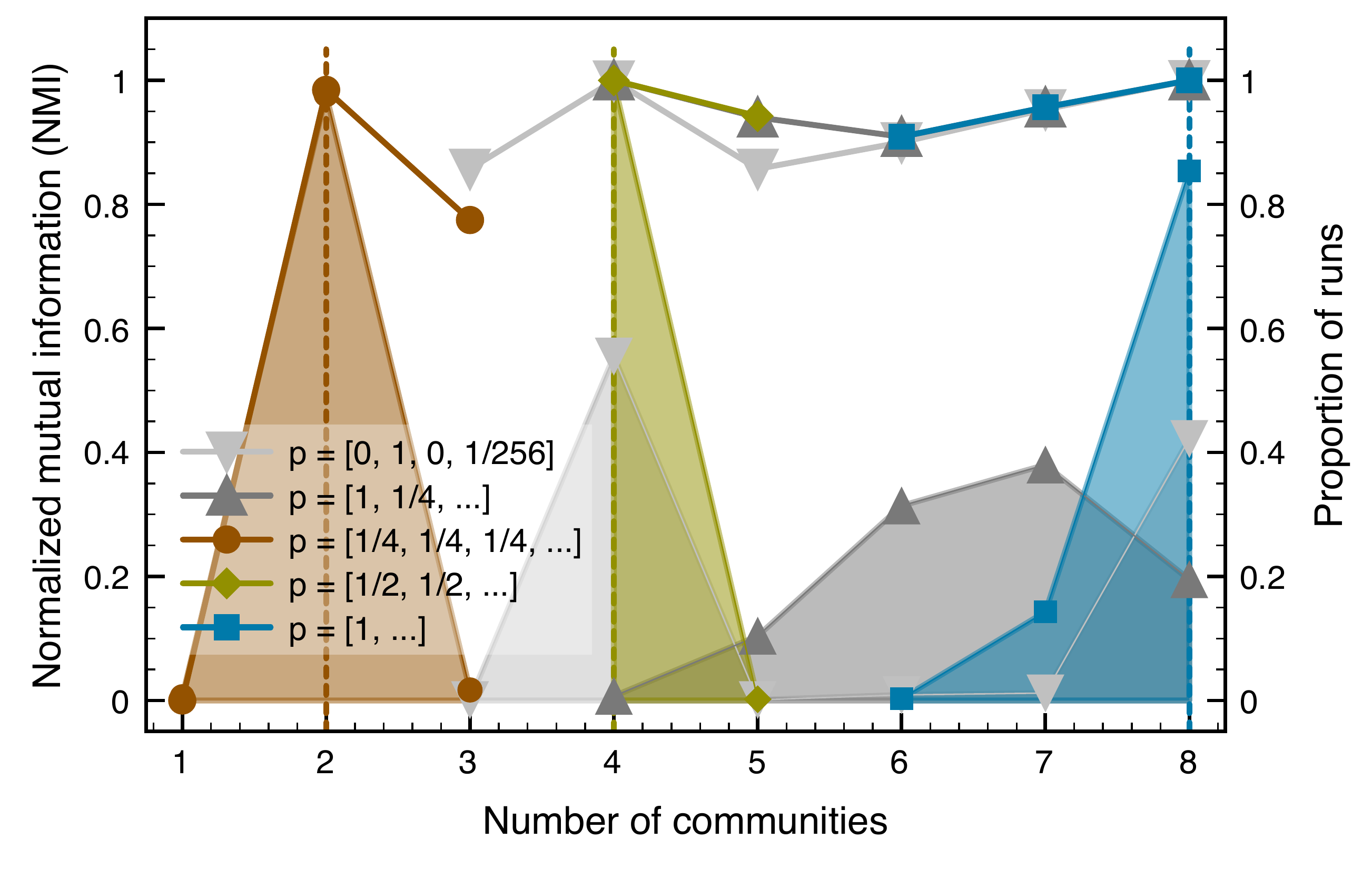}}
\caption{\label{fig_eval_HN}Analysis on HN benchmark networks~\cite{CMN08}. The values are estimates over $1000$ network realizations, while missing values of $p$ in the legend equal $\frac{1}{16}$. See also~text.}
\end{figure}

\paragraph{HN Benchmark.}Next, we also analyze the proposed algorithm on a class of benchmark networks with a hierarchical structure~\cite{CMN08}. In particular, networks are constructed according to a community dendrogram in~\figref{HN_dend}, where leafs correspond to eight modules of $16$ nodes, while each node $d$ of the dendrogram is also associated with a probability $p_d$, $p_d\in[0,1]$. The nodes of the network are linked with the probability associated with the lowest common ancestor in the community dendrogram. Varying the values of $p_d$ can infer (almost) arbitrary hierarchical structure of either link-density or link-pattern communities. However, due to simplicity, we associate each level of the nodes with the same probability $p_d$. Thus, denote $p=[p_1,p_2,p_3,p_4]$ to be the vector of respective probabilities for the nodes from the lowest to the highest level of the hierarchy, respectively.

The performance of \mpa~on five realizations of the above benchmark can be seen in~\figref{eval_HN}. Values of NMI were estimated such that each revealed partition was compared against (only) three intrinsic community structures---represented by dashed lines in~\figref{HN_dend}---and the best correspondence was reported. (Note that the results are thus actually rather pessimistic.) Observe that \mpa~can accurately reveal the planted structure in all five cases---see legend in~\figref{eval_HN_vals}---which further confirms the adequacy of the proposed community model. More precisely, in the first case, the intrinsic network structure results in a hierarchy of link-pattern communities, whereas, in the second case, communities are in fact defined on two levels of the designed hierarchy. In each of the last three cases, the communities corresponds to a single level of the hierarchy. Thus, \mpa~can indeed be employed for the detection of arbitrary community structure.

\begin{figure}[t]
\centering
\subfigure[\label{fig_eval_ER}Analysis on random graphs~\cite{ER59}]{
\includegraphics[width=0.475\textwidth]{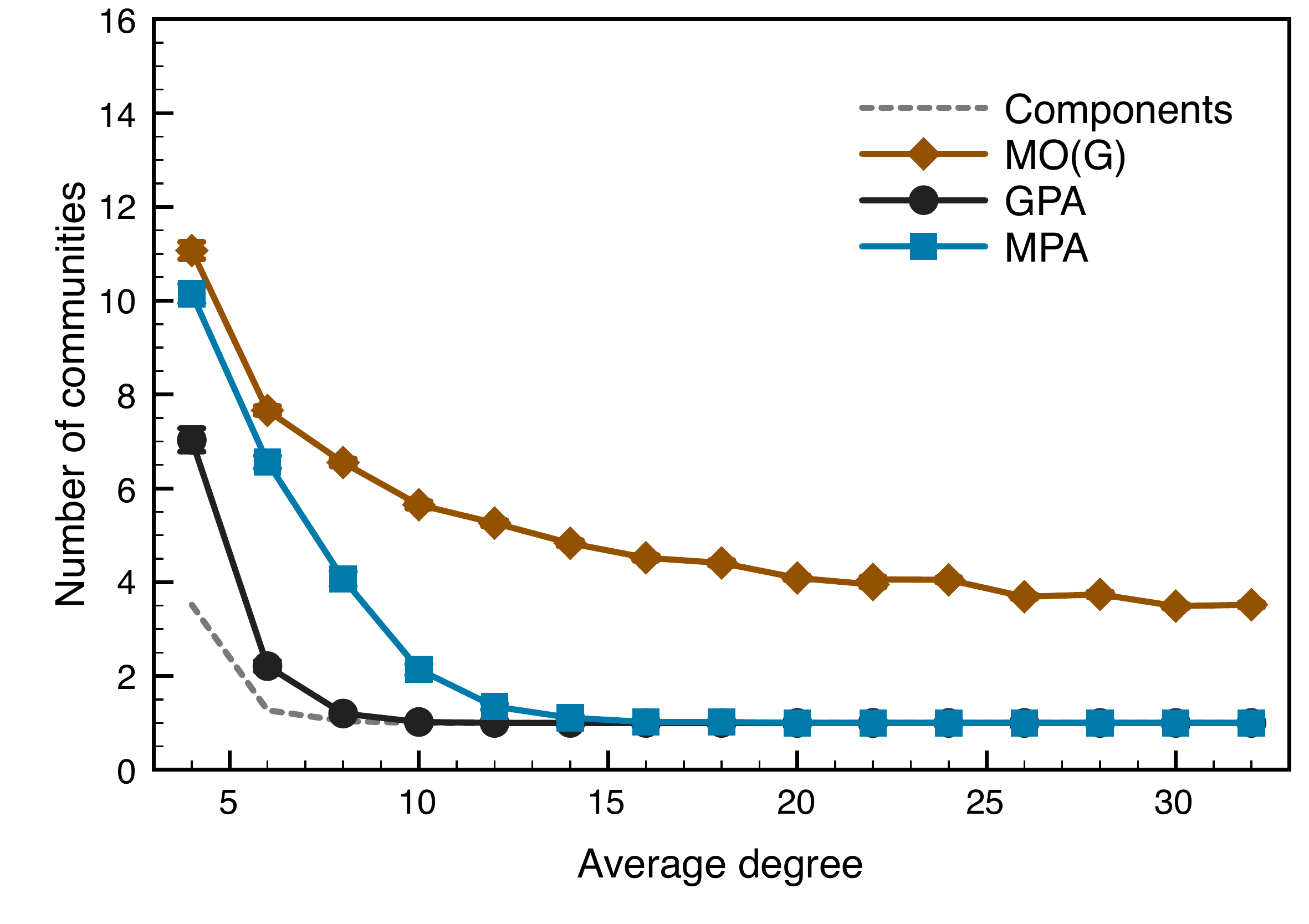}}
\subfigure[\label{fig_eval_RL}Analysis on a resolution limit test~\cite{FB07}]{
\includegraphics[width=0.475\textwidth]{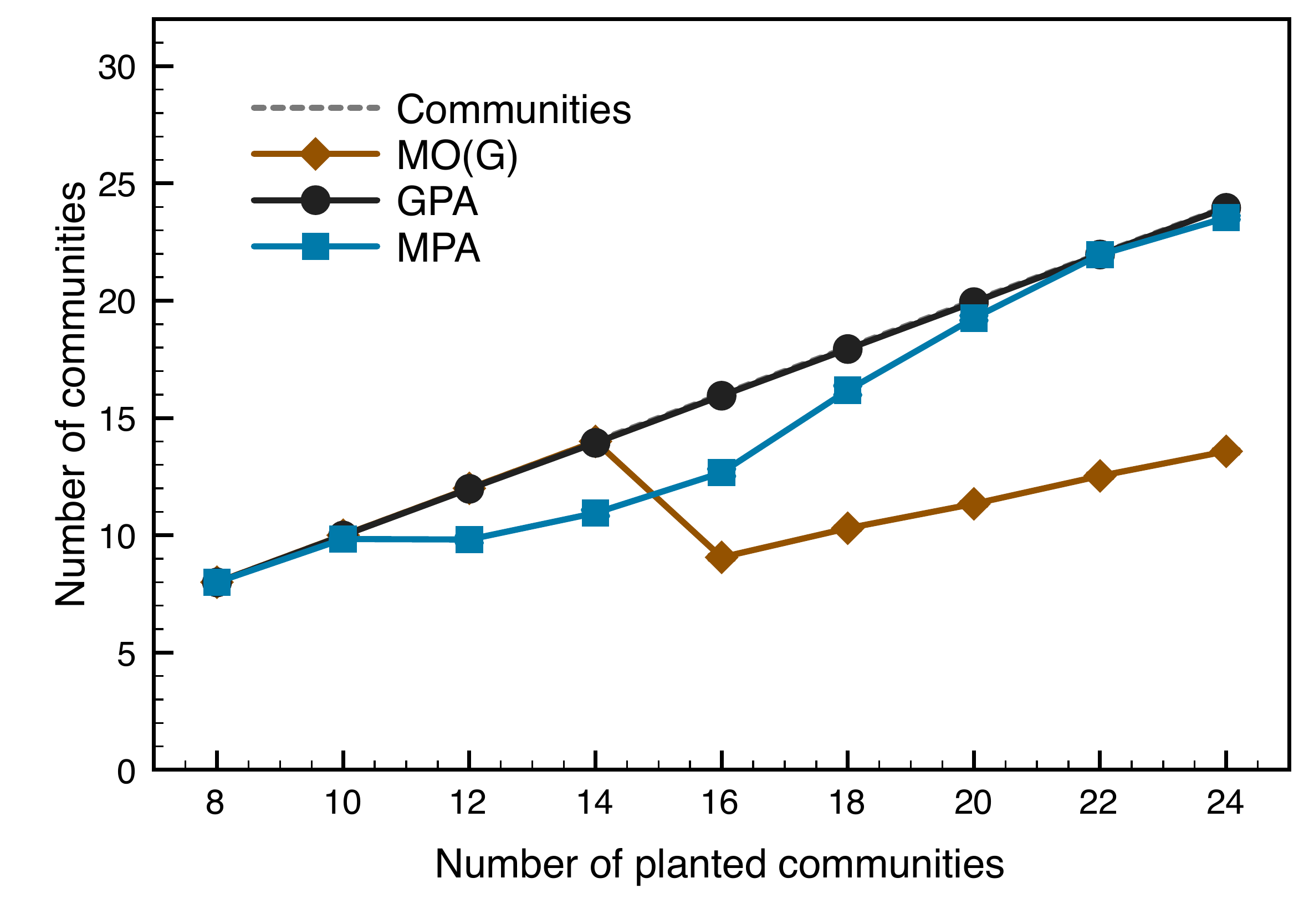}}
\caption{\label{fig_eval_ERRL}Analysis on (a) random graphs~\cite{ER59} and (b) resolution limit test networks~\cite{FB07}. The values are estimates over $100$ network realizations, while error bars are smaller than the symbol sizes.}
\end{figure}

\paragraph{Random Graphs.}We also apply the algorithms to Erd\"{o}s-R\'{e}nyi random graphs~\cite{ER59} that presumably have no community structure. We fix the number of nodes to $128$ and vary the average degree from $4$ to $32$. The results are shown in~\figref{eval_ER}. Note that, in contrast to \gmo, neither \mpa~nor \gpa~reports any community structure for these networks---all nodes are classified into a single community.  

\paragraph{Resolution Limit.}We further analyze the algorithms on a resolution limit~\cite{FB07}---existence of an intrinsic scale within the algorithm, below which the communities are no longer recognized---test benchmarks networks~\cite{FB07}. Hence, the networks consist of cliques with $4$ nodes that are linked into a ring. Results in~\figref{eval_RL} reveal that neither \mpa~nor \gpa~is seriously attributed to the resolution limit issues, whereas, the opposite holds for \gmo. Although some fluctuations are indeed observed for \mpa, these are not as severe as in the case of modularity~\cite{FB07}.

\begin{figure}[t]
\centering
\subfigure[\label{fig_eval_ZKC}Analysis on karate network~\cite{Zac77}]{
\includegraphics[width=0.475\textwidth]{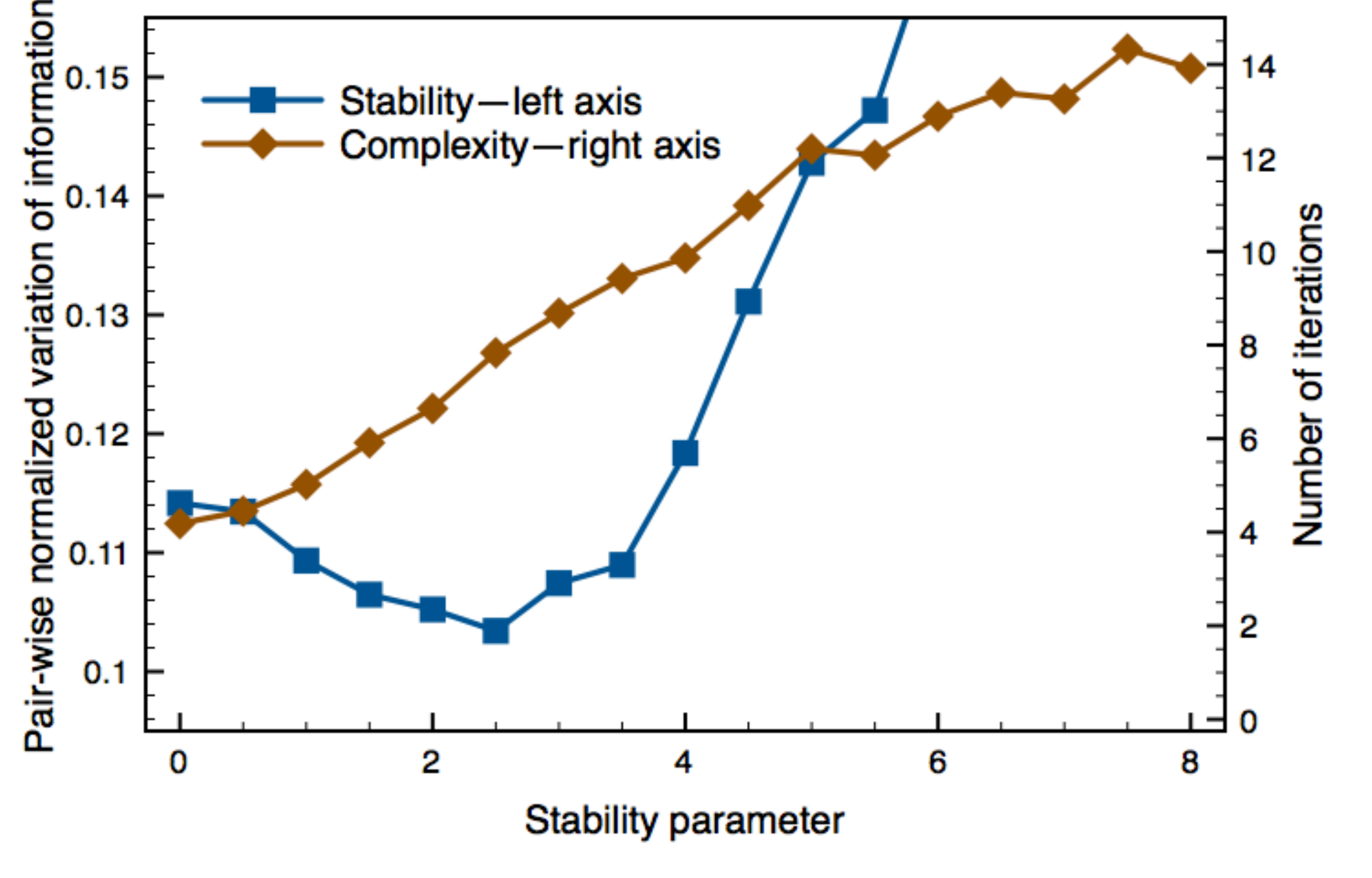}}
\subfigure[\label{fig_eval_AFL}Analysis on football network~\cite{GN02}]{
\includegraphics[width=0.475\textwidth]{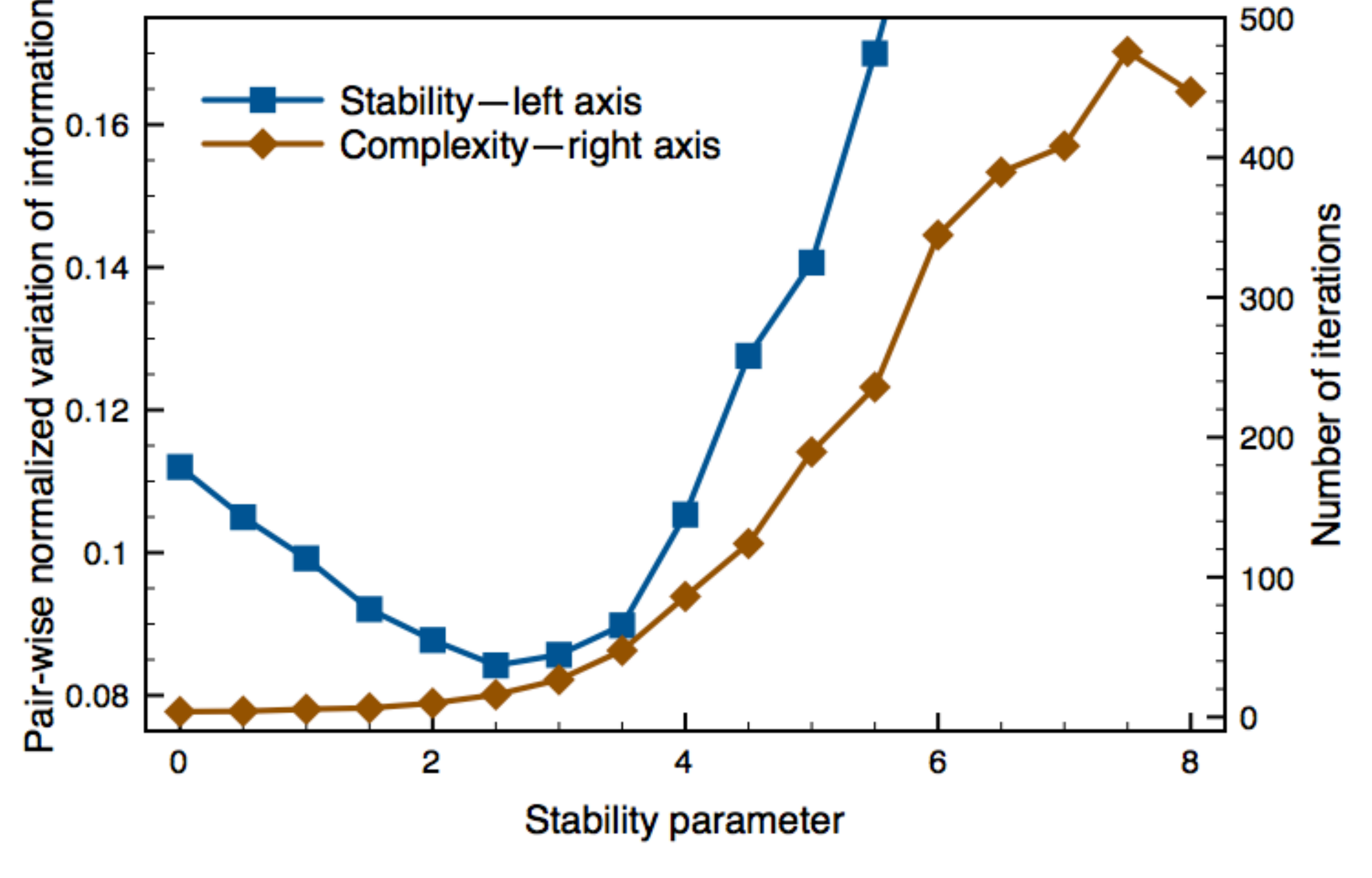}}
\subfigure[\label{fig_eval_SWC}Analysis on women network~\cite{DGG41}]{
\includegraphics[width=0.475\textwidth]{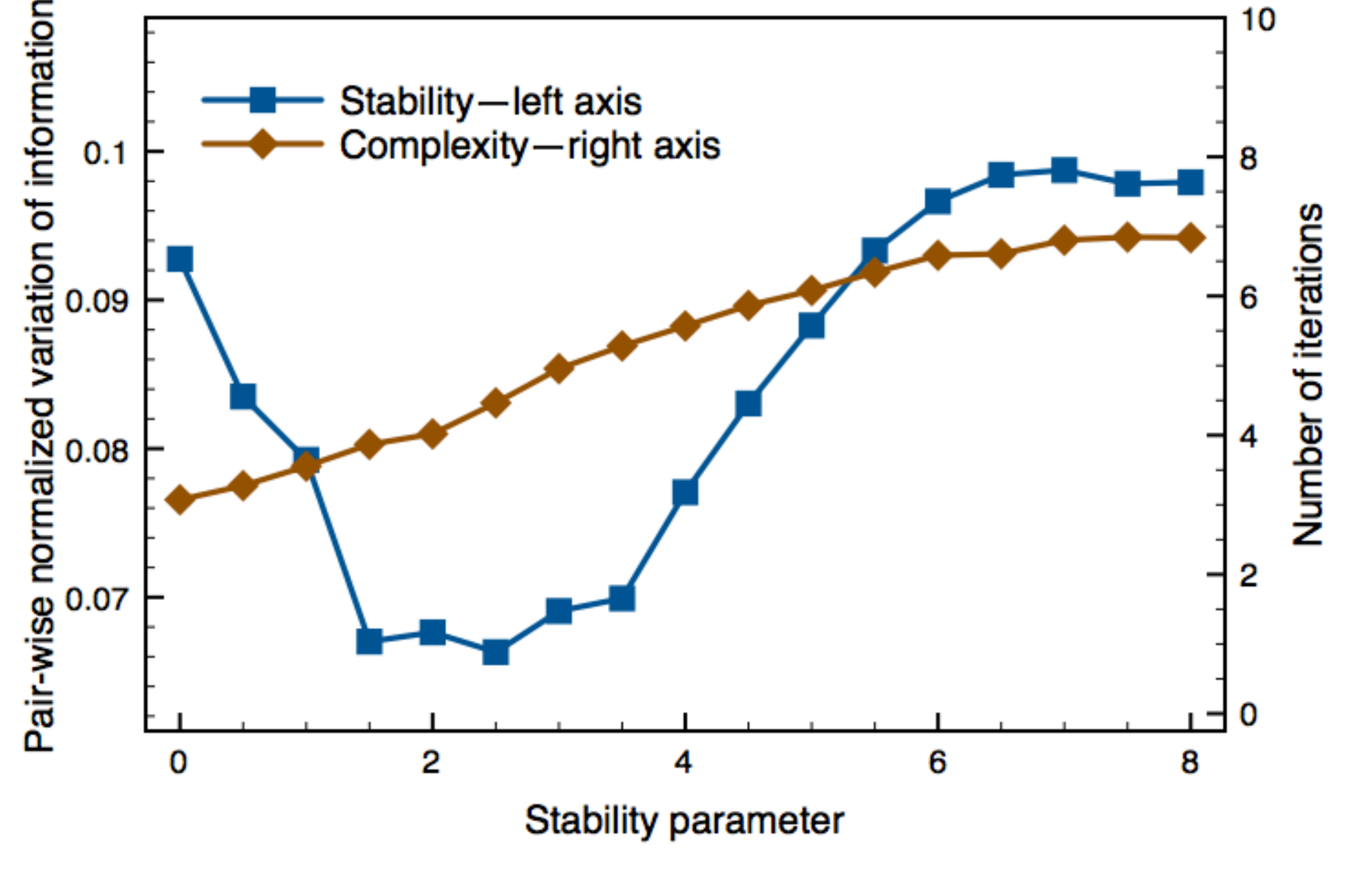}}
\caption{\label{fig_eval_ZAS}Analysis of stability and complexity of~\mpa~on three real-world networks from~\tblref{rw_cd}. The values are estimates over at least $100$ runs (note different scales).}
\end{figure}

\paragraph{Algorithm Stability.}As previously discussed, random label update orders severely hamper the stability of label propagation, and thus also the robustness of the revealed community structure~\cite{TK08}. Hence, balanced propagation~\cite{SB11b} is employed, yet this introduces two parameters $\lambda$ and $\mu$ (\secref{alg}). Value of $\lambda$ is intuitively fixed to $\frac{1}{2}$ (see~\eqref{b}), while parameter $\mu$ in fact controls the stability of the algorithm. In~\figref{eval_ZAS} we analyze \mpa~with respect to stability parameter $\mu$ on three real-world networks from~\tblref{rw_cd}. Plots show pair-wise distance between revealed community structures, and also the number of iterations for the algorithm to converge (note different scales). Observe that increasing $\mu$ improves the stability of \mpa~in all three networks, however, the number of iterations also increases. Furthermore, as one would expect, when $\mu$ exceeds a certain threshold, pair-wise distance between community structures notably increases---some number of nodes already gets completely disregarded due to propagation preferences close to $0$ (see~\eqref{b})---while the number of iterations can also increase substantially (see~\figref{eval_AFL}). The transition occurs at around $\mu\approx 4$ for these networks, thus, for the analysis throughout the paper, $\mu$ is set to $2$ (if not stated otherwise). It ought to be mentioned that balanced propagation can also improve community detection strength of the basic label propagation~\cite{SB11b} (see above).

\paragraph{Community Modeling.}For a comprehensive analysis, we also directly analyze the proposed community modeling strategy of \mpa~on SB benchmark networks with confusion degree set to $2$ (see above). In particular, we measure the average value of community parameter $\delta_c$ (\secref{alg}) for the nodes in each of the planted network communities. Results in~\figref{eval_SB_delta} show that, at least for these networks, values of community parameter $\delta_c$ clearly distinguish between link-density and link-pattern regime---average $\delta_c$ is close to $1$ and $0$ for the nodes in link-density and link-pattern communities, respectively. Note also that, due to lower average degree, values of $\delta_c$ are initially higher for the larger of the two link-pattern communities. However, before the algorithm converges---average number of iterations is shown by a horizontal line in~\figref{eval_SB_delta_vals}---community model in \mpa~infers the same average value of $\delta_c$ for both link-pattern communities. Note also that \gpa~cannot properly model communities planted in these networks (\figref{eval_SB}).

\paragraph{Computational Complexity.}Basic label propagation and its advances exhibit near linear time complexity $\mathrm{O}(|L|)$~\cite{RAK07}, where $|L|$ is the number of links in the network. In particular, the exact complexity was estimated to around $\mathrm{O}(|L|^{1.2})$~\cite{SB11d}. Similarly, the proposed model-based propagation \mpa~exhibits complexity near $\mathrm{O}(k|L|)$, where $k$ is the average degree in the network. Although a thorough empirical analysis is out of scope of this paper, based on the results in~\cite{SB11d} (and above), we estimate that \mpa~should scale up to networks with a million links---accessible on a standard desktop computer within an hour. 

\paragraph{Final Remarks.}The above analysis on different benchmark networks and random graphs indeed confirms that \mpa~can reveal arbitrary composites of either link-density or link-patter communities, as long as they are clearly depicted in the network's topology. Moreover, the proposed community modeling strategy also seems more adequate than the approach proposed by~\v{S}ubelj~and~Bajec~\cite{SB11c} for all networks considered. Further note that, although \mpa~is mostly outperformed by \emmm~on the benchmarks above, the latter should be attributed to the fact that \emmm~is advised about the number of communities. However, this currently cannot be properly estimated for large networks~\cite{KN11a}. Moreover, \mpa~also performs significantly better on real-world networks in~\secref{rwe}.

%	--------	--------	--------

\section{\label{sec_rwe}Real-world examples}
In the following we further employ the proposed algorithm for community detection in different unipartite and bipartite social networks---classical and fully link-pattern community detection, respectively---and also for a generalized community detection and predictive data clustering. All of the networks considered below are regarded as unweighted and undirected.

\paragraph{Community Detection}We first consider two classical networks for community detection---a network of social interations between members of a karate club analyzed by~Zachary~\cite{Zac77}, and a network of interplays in the $2000$ NCAA American football schedule proposed in~\cite{GN02}---and two well-known bipartite networks---a network of social collaborations between women in Natchez, Mississippi collected by~Davis~\cite{DGG41}, and a network of corporate interlocks in Scotland between $1904$ and $1905$ introduced in~\cite{SH80} (see~\tblref{rw_cd}). All these networks have known natural community structures that results from earlier studies (see also~\figref{comms}).

\begin{table}[t]
\caption{\label{tbl_rw_cd}Analysis on real-world networks subject to NMI estimated over $1000$ runs ($10$ runs for software networks). Corporate network is reduced to the largest component, while the known partition is also limited to $86$ corporate nodes---we thus set $\mu$ to $\frac{1}{2}$.}
\begin{center}
\begin{tabular}{crrccccc}
\hline
\multicolumn{1}{c}{\rule{0pt}{12pt}Network} & 
\multicolumn{1}{c}{\rule{0pt}{12pt}Nodes} & 
\multicolumn{1}{c}{\rule{0pt}{12pt}Links} & 
\multicolumn{1}{c}{\rule{0pt}{12pt}Comm.} & 
\multicolumn{1}{c}{\rule{0pt}{12pt}\gmo} & 
\multicolumn{1}{c}{\rule{0pt}{12pt}\gpa} & 
\multicolumn{1}{c}{\rule{0pt}{12pt}\emmm} & 
\multicolumn{1}{c}{\mpa}\\[2pt]
\hline\rule{0pt}{12pt}
Zachary's karate club~\cite{Zac77}  & $34$ & $78$ & $2$ & $0.6925$ & $0.7155$ & $0.7870$ & $\mathbf{0.8949}$ \\
American college football~\cite{GN02}  & $115$ & $616$ & $12$ & $0.7547$ & $0.8769$ & $0.8049$ & $\mathbf{0.8919}$ \\[2pt]
%\hline\rule{0pt}{12pt}
Davis's southern women~\cite{DGG41}  & $32$ & $89$ & $4$ & & $0.7338$ & $\mathbf{0.8332}$ & $0.8084$ \\ % NMI = 0.5475.
Scottish corpor. interlocks~\cite{SH80}  & $217$ & $348$ & $8$ & & $\mathbf{0.6634}$ & $0.5988$ & $0.6411$ \\[2pt] % NMI = 0.6301.
%\hline\rule{0pt}{12pt} 
%JUNG graph and network library~\cite{SB11a}  & $317$ & $719$ & $39$ & $\mathbf{0.6139}$ & $0.5663$ & $-$ & $0.5537$ \\ % 100 runs.
Java (\texttt{org} namespace)~\cite{SB11a}  & $709$ & $3571$ & $47$ & $0.5029$ & $\mathbf{0.5190}$ & $-$ & $\mathbf{0.5187}$ \\[2pt] 
Java (\texttt{javax} namespace)~\cite{SB11a}  & $1595$ & $5287$ & $107$ & $0.7048$ & $\mathbf{0.7369}$ & $-$ & $\mathbf{0.7386}$ \\[2pt] 
\hline
\end{tabular}
\end{center}
\end{table}

Propagation algorithms---\mpa~and~\gpa---most accurately reveal the true community structure for main of these networks (\tblref{rw_cd}), whereas, community modeling strategy of~\mpa~again seems more adequate than that of~\gpa. Note also that most values of NMI for \mpa~in~\tblref{rw_cd} are considerably high.

\begin{figure}[t]
\centering
\subfigure[Network adjacency matrix]{
\includegraphics[width=0.425\textwidth]{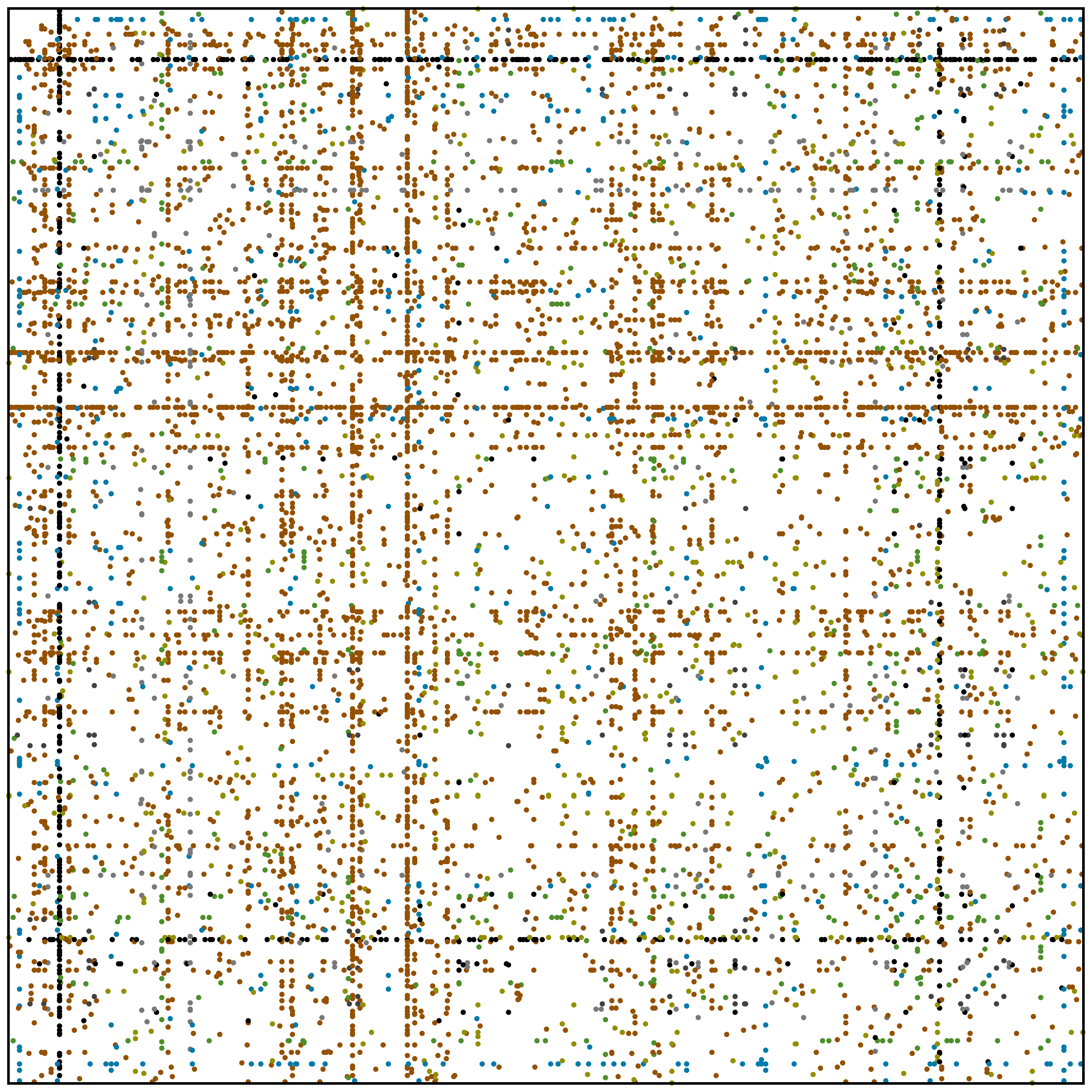}}
\subfigure[Blockmodel---reordered adj. mat.]{
\includegraphics[width=0.425\textwidth]{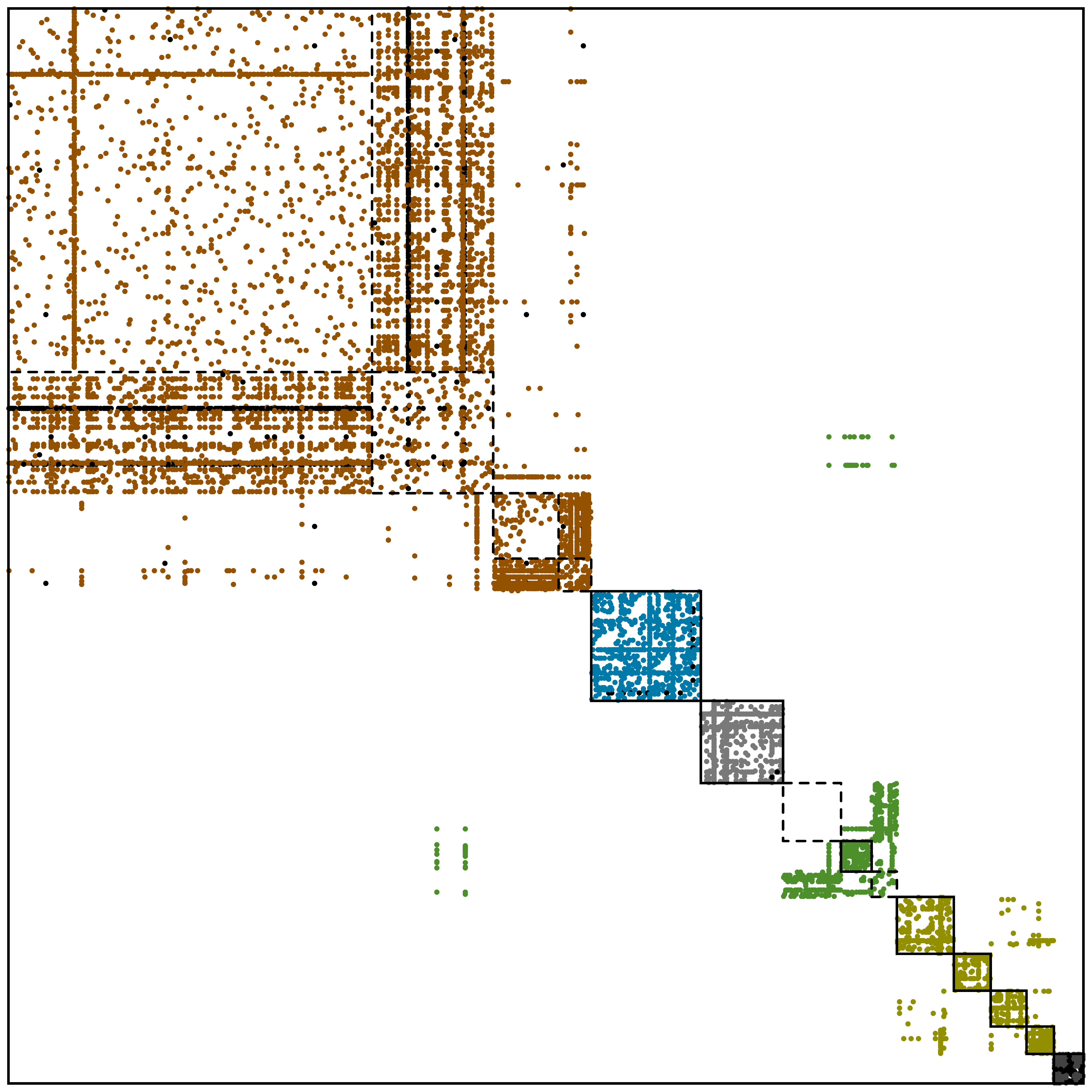}}
\caption{\label{fig_rw_javax}Community structure of Java software network revealed with \mpa~(b). Only communities with more than $24$ nodes are shown, still, the structure contains $1020$ nodes and $4184$ links. Link colors correspond to high-level software packages---{\color{brown}\texttt{javax.swing}}, {\color{blue}\texttt{javax.management}}, {\color{yellow}\texttt{javax.xml}}, {\color{green}\texttt{javax.print}}, {\color{lightgray}\texttt{javax.naming}}, {\color{darkgray}\texttt{javax.lang}} and {\color{black}other}---while each dot was enlarged five times for better visibility.} % Showing 14 of 92 communities. Hubs correspond to ComponentUI, JComponent and Accessible etc.
\end{figure}

Next, we also consider two software class dependency networks representing \texttt{org} and \texttt{javax} namespaces of Java language compiled in~\cite{SB11a}. Here, the natural community structure should coincide with respective software packages~\cite{SB11a}, while these are expected to conform with link-density and also link-pattern regime~\cite{SB11c}. Again, propagation algorithms most accurately extract the true network structures (\tblref{rw_cd}), whereas \emmm~fails completely. In~\figref{rw_javax} we also show the community structure of \texttt{javax} network revealed with \mpa~that obtains $\mathrm{NMI}=0.7431$. Observe how communities rather agree with high-level software packages, whereas, the majority of the links in the network is consistent with the revealed structure. Interestingly, some packages contain mainly link-pattern communities (e.g., {\color{brown}\texttt{javax.swing}}), while others are composed of only link-density communities (e.g., {\color{yellow}\texttt{javax.xml}}). 

\paragraph{Data Clustering}To apply community detection algorithm for data clustering, the respective dataset must first be represented by a network using some measure of similarity. According to~\cite{RDC11}, we adopt the inversed Chebyshev distance, with initial $[0,1]$-normalization. In order to obtain a sparse network, links must also be thresholded accordingly. (Due to simplicity, we consider only unweighted versions of the algorithms.) Note that the resulting network thus commonly decomposes into several connected components, however, community detection algorithm can still be employed to further partition these components (see~\tblref{rw_clus}).

\begin{table}[t]
\caption{\label{tbl_rw_clus}Analysis of data clustering on two real-world datasets subject to NMI and FCC, respectively (estimated over $100$ runs).} % http://archive.ics.uci.edu/ml/datasets.html
\begin{center}
\begin{tabular}{crcrcccc}
\hline
\multicolumn{1}{c}{\rule{0pt}{12pt}Dataset} & 
\multicolumn{1}{c}{\rule{0pt}{12pt}Items} & 
\multicolumn{1}{c}{\rule{0pt}{12pt}Classes} & 
\multicolumn{1}{c}{\rule{0pt}{12pt}Links} & 
\multicolumn{1}{c}{\rule{0pt}{12pt}Comp.} & 
\multicolumn{1}{c}{\rule{0pt}{12pt}\km} & 
\multicolumn{1}{c}{\rule{0pt}{12pt}\emmm} & 
\multicolumn{1}{c}{\mpa}\\[2pt]
\hline\rule{0pt}{12pt}
\multirow{2}{*}{Iris plants dataset~\cite{Fis36}} & \multirow{2}{*}{$150$}& \multirow{2}{*}{$3$} & \multirow{2}{*}{$2405$} & \multirow{2}{*}{$2$} & $\mathbf{0.8234}$ & $0.8113$ & $\mathbf{0.8264}$ \\
 & & & & & $0.8227$ & $0.8196$ & $\mathbf{0.8983}$ \\[2pt] % threshold = 0.15^-1.
%\hline\rule{0pt}{12pt}
\multirow{2}{*}{Ecoli protein dataset~\cite{HN96}} & \multirow{2}{*}{$336$} & \multirow{2}{*}{$8$} & \multirow{2}{*}{$14685$} & \multirow{2}{*}{$4$} & $0.5835$ & $0.0797$ & $\mathbf{0.6251}$ \\
 & & & & & $0.2530$ & $0.0277$ & $\mathbf{0.4164}$ \\[2pt] % threshold = 0.285-1.
\hline
\end{tabular}
\end{center}
\end{table}

We employ community detection to predict class variables of two famous datasets---Iris plants dataset introduced by~Fisher~\cite{Fis36}, and Ecoli protein localization sites dataset~\cite{HN96}. For comparison, in~\tblref{rw_clus} we also report the results for a classical partitional clustering algorithm K-Means~\cite{Mac67} (denoted \km). Observe that \mpa~obtains extremely promising results on these datasets, while it also significantly outperforms \emmm~and \km~that are both advised about the number of communities. Still, the results could be further improved in various ways. (Note that low NMI for~\emmm~on Ecoli dataset is not entirely evident.)

%	--------	--------	--------

\section{\label{sec_conc}Conclusions}
The paper proposes an enhanced community modeling strategy for a recently introduced general propagation algorithm~\cite{SB11c}. The resulting algorithm can detect arbitrary network modules---ranging from link-density communities to link-pattern communities---while, in contrast to most other approaches, it requires no apriori knowledge about the true structure (e.g., the number of communities). The algorithm was evaluated on various benchmark networks with planted partition, on random graphs and resolution limit test networks, where it is shown to be at least comparable to current state-of-the-art. Moreover, to demonstrate its generality, the algorithm was also employed for community detection in different unipartite and bipartite social networks, for generalized community detection and data clustering. The results imply that the proposed community model provides an adequate approximation of the real-world network structure, although, recent work suggests that network clustering and degree mixing could be even further utilized within the model~\cite{SV05,FFGP10,VGSTW10,RCS11}. The latter will be considered for future work. (For supporting website see~\url{http://lovro.lpt.fri.uni-lj.si/}.)

%	--------	--------	--------	--------	--------

%	-----------------
% 	ACKNOWLEDGMENTS
%	-----------------

\section*{Acknowledgments}
This work has been supported by the Slovene Research Agency ARRS within Research Program No. P2-0359.

%	-----------
% 	BIBLIOGRAPHY
%	-----------

\bibliographystyle{splncs03}
%\bibliography{refs}

\end{document}